\documentclass[aps,pra,twocolumn,10pt,superscriptaddress,amsfonts,amssymb,amsmath,floatfix,showkeys,nobibnotes,nofootinbib,a4paper]{revtex4-1}
\usepackage[T1]{fontenc}
\usepackage{lmodern}
\usepackage{graphicx}
\usepackage[colorlinks=true,allcolors=blue]{hyperref}

\makeatletter
\renewcommand{\p@subsection}{}
\renewcommand{\p@subsubsection}{}
\makeatother

\begin{document}

\title{Progress in cooling nanoelectronic devices to ultra-low temperatures}

\newcommand{\ulanc}{\affiliation{Department of Physics, Lancaster University, Lancaster, LA1~4YB, UK}}
\newcommand{\basel}{\affiliation{Department of Physics, University of Basel, CH-4056 Basel, Switzerland}}

\author{A.~T.~Jones}\ulanc
\author{C.~P.~Scheller}\basel
\author{J.~R.~Prance}\email[]{j.prance@lancaster.ac.uk}\ulanc
\author{Y.~B.~Kalyoncu}\basel
\author{D.~M.~Zumb\"uhl}\basel
\author{R.~P.~Haley}\ulanc

\begin{abstract}
Here we review recent progress in cooling micro/nanoelectronic devices significantly below $10\,\mathrm{mK}$. A number of groups worldwide are working to produce sub-millikelvin on-chip electron temperatures, motivated by the possibility of observing new physical effects and improving the performance of quantum technologies, sensors and metrological standards. The challenge is a longstanding one, with the lowest reported on-chip electron temperature having remained around $4\,\mathrm{mK}$ for more than 15 years. This is despite the fact that microkelvin temperatures have been accessible in bulk materials since the mid 20th century. In this review we describe progress made in the last five years using new cooling techniques. Developments have been driven by improvements in the understanding of nanoscale physics, material properties and heat flow in electronic devices at ultralow temperatures, and have involved collaboration between universities and institutes, physicists and engineers. We hope that this review will serve as a summary of the current state-of-the-art, and provide a roadmap for future developments. We focus on techniques that have shown, in experiment, the potential to reach sub-millikelvin electron temperatures. In particular, we focus on on-chip demagnetisation refrigeration. Multiple groups have used this technique to reach temperatures around $1\,\mathrm{mK}$, with a current lowest temperature below $0.5\,\mathrm{mK}$.
\end{abstract}

\keywords{Nanoelectronics; Ultra-low temperatures; Dilution refrigeration; Adiabatic nuclear demagnetization}

\maketitle

\section{I\lowercase{ntroduction}}
\label{sec:intro}

Millikelvin electronic measurements of micro/nanoscale devices and materials are used in a wide range of fields; from quantum technology, materials science and metrology to observational astrophysics and dark matter searches. In some cases, physical effects emerge at low temperature that provide a new and useful electronic behaviour, such as superconductivity or conductance quantisation. In other cases, low temperatures provide a ``quiet'' environment that can, for example, improve the signal-to-noise ratio of sensitive detectors or increase the coherence time of qubits. Regardless of the goal, or the refrigeration technology used, it remains challenging to cool the conduction electrons in a nanoscale device or material significantly below $10\,\mathrm{mK}$. As the temperature drops, the thermal coupling between conduction electrons and the host lattice weakens and the heat capacity of the electronic system falls. This makes the electron temperature more sensitive to parasitic heating. In a nanoscale structure, where the physical size already limits the electronic heat capacity, it is very challenging to maintain low electron temperatures against the incoming heat from electromagnetic radiation, eddy-current heating, nearby hot insulators, and the electronic connections needed for measurement. This review outlines the current progress in cooling nanoelectronic systems below $10\,\mathrm{mK}$, and the potential for new techniques to reach electron temperatures deep in the microkelvin regime.

The ability to access low-millikelvin or microkelvin temperatures in nanoelectronic structures brings the exciting possibility of unexpected discoveries in a new regime. But there are also immediate goals that motivate much of the work we discuss here. Low electron temperatures are needed to observe new predicted electronic phases, including exotic quantum Hall states~\cite{Chesi2008,Nayak2008,Stern2010,Pan2015,Samkharadze2015}, topological insulators~\cite{Hasan2010}, collective electron-nuclear spin states~\cite{Simon2007,Simon2008,Braunecker2013,Scheller2014a}, insulating ground states in 2D systems~\cite{Huang2007,Knighton2018} and superconductivity in some materials~\cite{Schuberth2016}. In established applications, lower electron temperatures may improve coherence times of semiconductor and superconducting qubits~\cite{Hanson2007,Clarke2008,Devoret2013} and hybrid Majorana devices~\cite{Lutchyn2010,Oreg2010,Alicea2010}, as well as reducing error mechanisms in metrological standards such as charge pumps~\cite{Pekola2013,Nakamura2014} and quantum Hall resistance standards~\cite{Matthews2005}.

This review focuses on cooling techniques that we know to have successfully produced on-chip electron temperatures significantly below $10\,\mathrm{mK}$ in experiment. We will not discuss emerging refrigeration techniques, such as micro/nanoscale electronic coolers, that may be able to reach ultralow temperatures but have not yet done so in experiment. More information on micrometer-scale refrigeration can be found in recent reviews such as~\cite{Giazotto2006,Muhonen2012}. We will also not discuss ultralow temperature thermometry in detail, although this is obviously an important and relevant topic. Information about the current state of metrology in ultralow temperature thermometry can be found in~\cite{Engert2016}. More information about techniques that are particularly relevant to micro/nanoelectronic devices at ultralow temperatures can be found, for example, in~\cite{Casey2014,Shibahara2016,Rothfuss2016} for noise thermometry,~\cite{Casparis2012,Meschke2016,Bradley2016,Hahtela2016} for Coulomb blockade thermometry and~\cite{Scheller2014,Iftikhar2016,Nicoli2019} for quantum dot-based thermometry. Almost all of the work discussed below makes use of one or more of these thermometry techniques.

\section{C\lowercase{ooling techniques and heat flow in nanoelectronic devices}}
\label{sec:heatflow}

When trying to cool micro/nanoelectronic devices to ultralow temperatures, experimentalists are faced with several unfavourable physical scaling laws: the heat capacity of the conduction electrons falls with temperature, as does their thermal coupling to other electronic systems and to phonons in the host lattice. To achieve an electron temperature $T_\mathrm{e}$ that is close to the base temperature of the surrounding environment, parasitic heat leaks into the conduction electrons need to be carefully managed. What this means quantitatively depends on the details of each sample and how it is coupled to its local environment; however, the scaling laws, discussed in more detail below, demonstrate the extent of the challenge of moving to lower temperatures. As a trivial example, consider a device that has been well-thermalised to a refrigerator operating at $10\,\mathrm{mK}$ by cooling through bond wires and keeping the total parasitic heat leak below $1\,\mathrm{fW}$. The same device could require a total heat leak below $10\,\mathrm{aW}$ to stay similarly well-thermalised to a refrigerator operating at $1\,\mathrm{mK}$. In this section, we outline a general thermal model for an on-chip conductor at low temperatures and use this model to illustrate the various cooling techniques that can be employed to reach on-chip electron temperatures below $10\,\mathrm{mK}$.

\subsection{Thermal model}

\begin{figure*}
	\includegraphics[width=1.4\columnwidth]{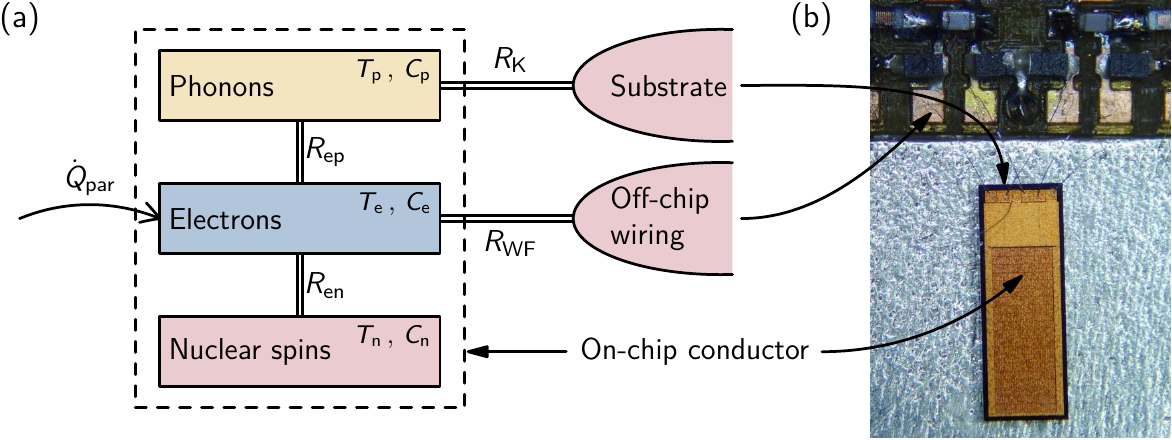}
	\caption{Thermal model of an on-chip conducting material at low temperature. The on-chip conductor [dashed box in (a)] contains three thermal subsystems: phonons, conduction electrons and nuclear spins, with heat capacities $C_\mathrm{p}$, $C_\mathrm{e}$, $C_\mathrm{n}$ and temperatures $T_\mathrm{p}$, $T_\mathrm{e}$, $T_\mathrm{n}$ respectively. Heat flow between the subsystems is determined by temperature differences and the thermal resistances $R_\mathrm{ep}$ and $R_\mathrm{en}$. The conductor sits on an insulating substrate, which is assumed to be macroscopic and in thermal equilibrium with the base temperature of an external refrigerator. The thermal resistance between the conductor and the substrate is the phonon boundary (Kapitza) resistance $R_\mathrm{K}$. The conductor is electrically connected to off-chip wiring, which is also assumed to be well-thermalised with the external refrigerator. The thermal resistance $R_\mathrm{WF}$ between on-chip electrons and electrons in the wiring is determined by the electrical resistance of the connection. (b) illustrates the location of each component in an optical image of a typical device on a low-temperature sample mount.}
	\label{fig:thermal_model}
\end{figure*}

The thermal model is outlined in Fig.~\ref{fig:thermal_model}. It shows several channels that are available to remove heat from the conduction electrons in an on-chip material. The thermal resistances of the channels are temperature dependent and so the optimal way to cool the electrons will also change with temperature. The details of the thermal resistance for each cooling channel may be different in different samples, but the illustration in Fig.~\ref{fig:thermal_model} is often a useful approximation and could apply to, for example, conduction electrons in a semiconductor nanostructure or in a thin-film metal circuit. In the following discussion we use the simple example of a metal conductor on the surface of an insulating substrate.

In the first instance, conduction electrons in an on-chip material are coupled to phonons and spins in the same material. In many commonly-used metals and semiconductors, the low-temperature thermal coupling between conduction electrons at temperature $T_\mathrm{e}$ and phonons at temperature $T_\mathrm{p}$ is given by the heat flow
\begin{equation}
	\dot{Q}_\mathrm{ep} = \Sigma V\left(T_\mathrm{e}^5 - T_\mathrm{p}^5\right)\;,
	\label{equ:Q_ep}
\end{equation}
where $\Sigma$ is a material-dependent coupling constant and $V$ is the volume of the material. Note that the exponent of the temperatures in this equation is commonly accepted to be $5$ in many materials~\cite{Pobell2007,Wellstood1994,Echternach1992}, however in some systems, particularly those confined to fewer dimensions, it has been observed to take other values $2 < n \leq 5$~\cite{Casparis2012,Bradley2016}. Typical values of $\Sigma$ measured below $1\,\mathrm{K}$ range from $\sim 0.01 \times 10^9\,\mathrm{W\,m^{-3}\,K^{-5}}$ in doped semiconductors to $\sim 1 \times 10^9\,\mathrm{W\,m^{-3}\,K^{-5}}$ in metals~\cite{Giazotto2006}.

If the on-chip material contains spinful nuclei, heat will flow between the nuclear spin bath and conduction electrons through spin-lattice relaxation. In the limit of small nuclear Zeeman splitting ($g_n \mu_n B \ll k_\mathrm{B} T$, where $g_n$ is the g-factor, $\mu_n$ is the nuclear magneton and $B$ the magnetic field), the spin-lattice relaxation rate $\tau_1^{-1}$ is proportional to $T_\mathrm{e}$ and characterised by the Korringa constant $\kappa = \tau_1 T_\mathrm{e}$~\cite{Korringa1950}. Also in this limit, the heat capacity of the nuclei is above the Schottky anomaly and follows $C_\mathrm{n} \propto B^2/T^2_\mathrm{n}$. The thermal coupling between conduction electrons and the nuclear spin bath at temperature $T_\mathrm{n}$ is then given by the heat flow~\cite{Huiskamp1973,Lounasmaa1974}
\begin{equation}
	\dot{Q}_\mathrm{en} = \frac{\lambda_n n}{\mu_0 \kappa} \frac{B^2}{T_\mathrm{n}} \left(T_\mathrm{e} - T_\mathrm{n}\right) \;,
	\label{equ:Q_en}
\end{equation}
where $\lambda_n$ is the molar nuclear Curie constant of the material, $n$ is the number of moles and $\mu_0$ is the permeability of free space. Equation~\ref{equ:Q_en} has been experimentally verified in a broad range of metals and semiconductors~\cite{Pobell2007} and we will assume that it is valid in the following discussion. However, it should be noted that deviations from the Korringa law, which invalidate Equ.~\ref{equ:Q_en}, have been observed in some metallically-doped semiconductors below $10\,\mathrm{K}$ in the disordered, interacting regime~\cite{Koelbl2012}, semiconductors doped close to the metal-insulator transition~\cite{Paalanen1985} and Kondo metals~\cite{Roshen1982}.

The thermal model in Fig.~\ref{fig:thermal_model} shows an on-chip material that contains a thermal bath of nuclear spins. The same basic model could also apply to a material that contains paramagnetic impurities. In this case, the nuclear spin bath is replaced by a bath of electron spins bound to impurities or dopants. The thermal resistance between these spins and the conduction electrons will be determined by the spin relaxation time. The heat capacity of the spin bath is likely to include a Schottky anomaly in the millikelvin temperature range~\cite{Vlasov2017}.

While the heat flows described by Equ.~\ref{equ:Q_ep} and \ref{equ:Q_en} redistribute energy between the thermal subsystems of an on-chip material, the thermal resistances $R_\mathrm{WF}$ and $R_\mathrm{K}$ determine how well the material is coupled to the outside environment. The thermal resistance $R_\mathrm{WF}$ represents electronic heat conduction through the electrical connections to a device. It is related to the electrical resistance $R$ of the connection via the Wiedemann--Franz law
\begin{equation}
	R_\mathrm{WF} = \frac{3}{\pi^2} \left(\frac{e}{k_\mathrm{B}}\right)^2 \frac{R}{(T_\mathrm{e} + T)/2} \;,
	\label{equ:R_WF}
\end{equation}
where $T$ is the temperature of the outside environment. In practice, the value of $R$ can be chosen across a wide range. The resistance of a single gold bond wire, including contact resistance, can be less than $10\,\mathrm{m\Omega}$ at low temperatures~\cite{Vliet2018}. On the other hand, the electrical resistance can easily be increased above $10\,\mathrm{k\Omega}$ by including on-chip thin-film resistors or tunnel junctions~\cite{Schicke2001,Jhabvala2002,Satrapinski2011,Roschier2012}.

The final component of the thermal model is the phonon boundary (Kapitza) resistance $R_\mathrm{K}$, which is typically between the on-chip conductor and the substrate. The value of $R_\mathrm{K}$ depends on the substrate material and the on-chip material, as well as the microscopic properties of the interface~\cite{Swartz1989}. The boundary resistance roughly scales as $R_\mathrm{K} \propto T^{-3}$ with a prefactor that depends on the acoustic mismatch between the two materials, the strength of scattering at the interface and the area $A$ of the interface. For common metals (including Al, Cu, Au, In) on insulating substrates (Sapphire, Quartz, Si) expected values are $A R_\mathrm{K}T^3 \sim 10^{-2}\,\mathrm{K^4\,m^2\,W^{-1}}$~\cite{Swartz1989}. Because it is difficult to control the quality of interfaces in experiment, a precise prediction of $R_\mathrm{K}$ is rarely possible. However, at ultralow temperatures it is common to find $R_\mathrm{ep} \gg R_\mathrm{K}$ and therefore cooling of the conduction electrons through phonon channels is not limited by $R_\mathrm{K}$. In some samples, for example a semiconductor 2D electron gas, the conduction electrons are inside the substrate material and couple directly to the substrate phonons. In this case, $R_\mathrm{K}$ may be omitted from the thermal model or it may be used to represent the boundary resistance between the substrate and its support.

\begin{figure*}
	\includegraphics[width=1.5\columnwidth]{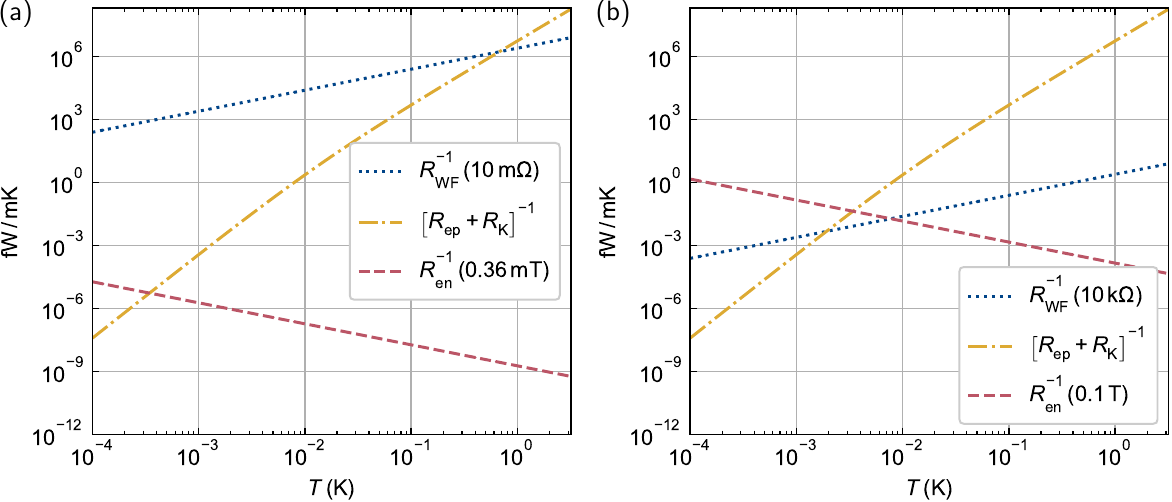}
	\caption{Predicted thermal conductances for the model shown in Fig.~\ref{fig:thermal_model} in two example situations. In both, the on-chip conductor is a copper film of size $205\,\mathrm{\mu m} \times 38.5\,\mathrm{\mu m} \times 5\,\mathrm{\mu m}$ (similar to the device in~\cite{Bradley2017}). Its substrate is a silicon chip, which is assumed to be well-thermalised to the external refrigerator at temperature $T$. In (a), the on-chip electrons are electrically connected to well-thermalised external wiring through a low-resistance ($10\,\mathrm{m\Omega}$) bond wire. This path provides the strongest thermal connection to the electrons for $T \ll 1\,\mathrm{K}$. No external magnetic field is applied and the internal magnetic field is assumed to be $0.36\,\mathrm{mT}$, the effective dipolar field in copper. In (b), the resistance of the electrical connection is $10\,\mathrm{k\Omega}$ and a magnetic field of $0.1\,\mathrm{T}$ is applied. As a result, coupling between the on-chip electrons and the external refrigerator is much weaker for $T \ll 1\,\mathrm{K}$ and, below a few millikelvin, the nuclear spin bath in copper becomes strongly coupled to the electrons.}
	\label{fig:thermal_resistances}
\end{figure*}

All of the thermal channels shown in Fig.~\ref{fig:thermal_model} have temperature-dependent thermal resistances. Figure~\ref{fig:thermal_resistances} shows predicted values of the corresponding thermal conductances for two example situations. In the first example, shown in Fig.~\ref{fig:thermal_resistances}(a), a thick ($\sim \mathrm{\mu m}$) on-chip copper film has a low-resistance electrical connection to some off-chip wiring. Both the external wiring and substrate chip are assumed to be macroscopic and well-thermalised with the refrigerator. Above $\sim 1\,\mathrm{K}$, the on-chip conduction electrons are primarily coupled to the refrigerator through phonons. At lower temperatures, the phonon channel closes rapidly and the bond wire provides the strongest thermal connection to the refrigerator. At temperatures $\ll 1\,\mathrm{K}$, cooling of the on-chip electrons will mostly happen through the bond wire, with a base electron temperature determined by the parasitic heating $\dot{Q}_\mathrm{par}$ and the thermal resistance $R_\mathrm{WF}$. The second example, shown in Fig.~\ref{fig:thermal_resistances}(b), is a similar system but with a $0.1\,\mathrm{T}$ magnetic field present and $10\,\mathrm{k\Omega}$ electrical resistance added between the external conductor and the on-chip copper. In this case, the magnetic field increases the thermal coupling between the conduction electrons and the spin-3/2 nuclei in the copper, and the electrical resistance is large enough to thermally isolate the conduction electrons from the off-chip wiring across the whole temperature range. This example shows that, for some devices, the conduction electrons can be most strongly coupled to other on-chip thermal subsystems at ultralow temperatures. It also demonstrates the challenge of cooling high-impedance devices such as single-electron transistors (SETs) or semiconductors with resistive ohmic contacts. Comparing the two examples in Fig.~\ref{fig:thermal_resistances}, the parasitic heating would need to be $10^6$ times smaller for the high-impedance case to to reach the same electron temperature as the low impedance case.

\begin{figure*}
	\includegraphics[width=1.5\columnwidth]{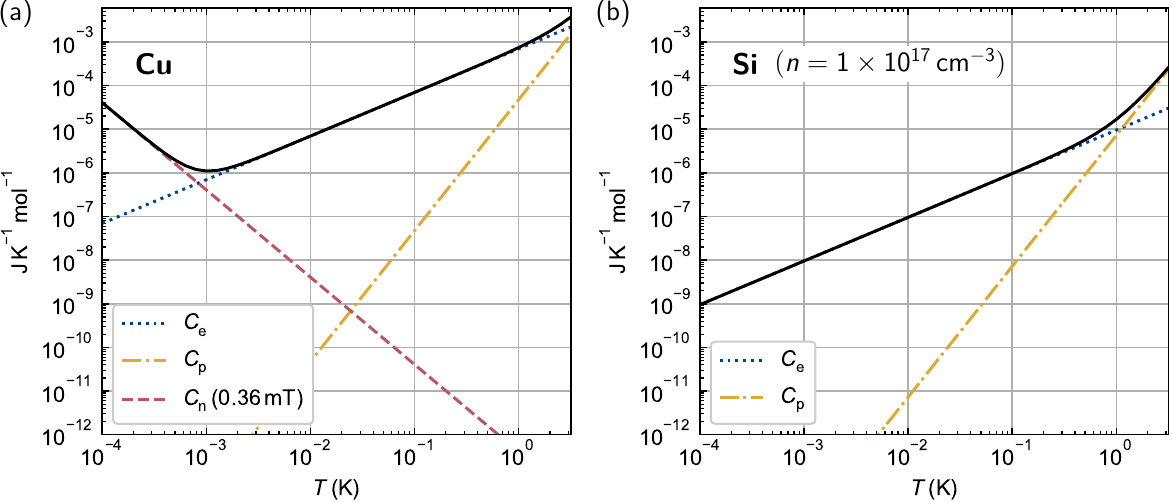}
	\caption{Molar heat capacities of copper and silicon at low temperatures. (a) Total heat capacity (solid line) of copper, which is the sum of contributions from conduction electrons $C_\mathrm{e}$, phonons $C_\mathrm{p}$ and nuclear spins $C_\mathrm{n}$. (b) Total heat capacity (solid line) of undoped silicon in zero applied magnetic field with a free electron concentration of $1\times 10^{17}\,\mathrm{cm^{-3}}$. This could be, for example, silicon in the channel of an accumulation-mode FET. In both materials, $C_\mathrm{p}$ is insignificant for $T \ll 1\,\mathrm{K}$. Even in zero applied magnetic field, the total heat capacity of copper is dominated by $C_\mathrm{n}$ for $T \ll 1\,\mathrm{mK}$. The contribution from $C_\mathrm{n}$ grows as the square of applied magnetic field.}
	\label{fig:heat_capacities}
\end{figure*}

The steady-state electron temperature in the thermal model is only determined by thermal resistances, the amount of parasitic heating and the temperature of the cold reservoir (refrigerator). However, the heat capacities of the various subsystems are needed to understand any dynamic behaviour. Figure~\ref{fig:heat_capacities} shows how the heat capacities of two example materials vary with temperature between $3\,\mathrm{K}$ and $100\,\mathrm{\mu K}$. The heat capacity of the conduction electrons falls linearly with temperature, making the instantaneous on-chip electron temperature more sensitive to intermittent sources of heat. In the case of undoped silicon, shown in Fig.~\ref{fig:heat_capacities}(b), its total heat capacity drops all the way down to $100\,\mathrm{\mu K}$, where it reaches a value $\sim 10^{10}$ times smaller than the room temperature phonon heat capacity. The situation is different in copper, as shown in Fig.~\ref{fig:heat_capacities}(a), because its total heat capacity is boosted below $\sim 1\,\mathrm{mK}$ by the presence of a nuclear spin bath. The nuclear heat capacity grows with applied magnetic field, moving the crossover to higher temperatures. At ultralow temperatures, the heat capacity of copper will be dominated by this contribution, even with zero applied magnetic field due to the internal magnetic field $b=0.36\,\mathrm{mT}$~\cite{Pobell2007}. The heat capacity of the nuclear spin bath can be used to stabilise the electron temperature [since $R_\mathrm{en} \ll R_\mathrm{ep}$, as shown in Fig.~\ref{fig:thermal_resistances}(b)] and even to cool the electrons through demagnetisation refrigeration, as discussed in later sections.

\subsection{Parasitic heat leaks and electrical filtering}

Eliminating parasitic heat leaks is one of the major challenges in cooling nano-electronic devices down to ultra low temperatures. Material heat release, microwave radiation from higher temperature stages of the dilution refrigerator as well as RF and low frequency noise coupling to the sample though its electrical leads are well known sources of parasitic heating. The main countermeasures include installation of radiation shields, thermal anchoring of the sample leads at multiple temperature stages of the dilution refrigerator, elimination of ground loops and, in particular, intensive microwave filtering. Various different filtering approaches have been proposed in the literature, a summary of which can be found in~\cite{Bladh2003}. These designs include metal powder filters~\cite{Scheller2014,Lukashenko2008,Milliken2007,Fukushima1997}, micro-fabricated filters~\cite{Sueur2006,Vion1995,Courtois1995,Longobardi2013}, thermocoax cables~\cite{Zorin1995,Glattli1997}, copper `tape worm' filters~\cite{Bluhm2008,Spietz2006a}, thin film filters~\cite{Jin1997} and lossy transmission lines~\cite{Slichter2009}.

\begin{figure}
	\centering
	\includegraphics[width=0.8\columnwidth]{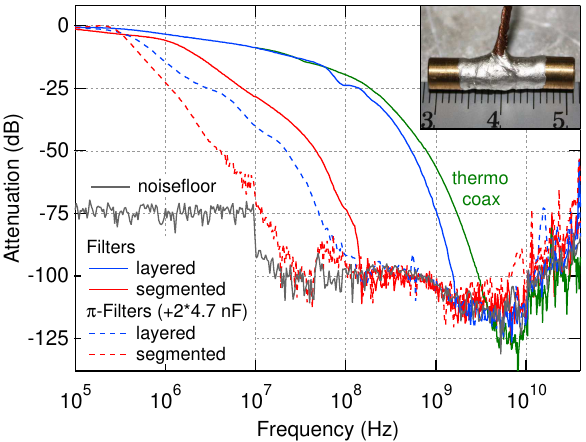}
	\caption{Room temperature attenuation characteristics of a $1.5\,\mathrm{m}$ long thermocoax cable (green) and different silver epoxy microwave filters. Blue and red represent layered and segmented filters, respectively, where the segmented filters have reduced parasitic capacitance. For the dashed characteristics, a $4.7\,\mathrm{nF}$ discoidal capacitor from Pacific Aerospace~\cite{pacificaerospace} was added to both filter ends. A picture of a silver epoxy microwave filter and centimetre scale bar is shown in the inset. This figure was taken from~\cite{Scheller2014}.}
	\label{fig:filters}
\end{figure}

Depending on the application, specific filtering designs may have advantages over others, for example the use of $50\,\mathrm{\Omega}$ characteristic filters~\cite{Milliken2007} when impedance matching is crucial, or dissipative cryogenic filters with zero dc resistance~\cite{Bluhm2008} for low impedance devices. Thermocoax cables~\cite{Zorin1995,Glattli1997} provide very strong attenuation in the microwave and THz regime, and can be used as signal wires from room temperature down to the mixing chamber. However, thermalisation of the inner conductor carrying the measurement signal is rather challenging, and filtering in the MHz regime is not as effective. Therefore a combination of thermocoax cables for the high frequency range and, for example, low cut-off frequency silver-epoxy microwave filters with improved thermalisation~\cite{Scheller2014}, as used for microkelvin experiments at the University of Basel (see Fig.~\ref{fig:filters}), ensures good filtering throughout the relevant frequency range and optimal thermal anchoring.

\subsection{Cooling techniques}

The combination of higher thermal resistances and lower electronic heat capacity makes it difficult to reach on-chip electron temperatures below $10\,\mathrm{mK}$ using standard experimental techniques in a dilution refrigerator. The most common approach, which works well at higher temperatures, is to ensure that the substrate and external wiring are well thermalised through solid contact with the coldest stage of the refrigerator, and to reduce parasitic heating and dissipation in the device as much as possible. Often, the latter requires careful filtering of electrical noise in the incoming wiring. For successful examples demonstrating $6\,\mathrm{mK} \lesssim T_\mathrm{e} \lesssim 10\,\mathrm{mK}$, see~\cite{Chung2003,Spietz2006,Potok2007,Bid2009,McClure2012,Scheller2014,Iftikhar2016}. Reaching significantly lower on-chip electron temperatures requires a different approach to thermalising the sample and different refrigeration technology, since even the best dilution refrigerators are limited to temperatures above $1\,\mathrm{mK}$~\cite{Cousins1999}.

Demagnetisation cooling is, at present, the most widely used technique for cooling bulk materials below the base temperature of a dilution refrigerator. It is used in low temperature laboratories~\cite{Pickett2000} and has been applied, although much less widely, to cool micro/nanoelectronic devices. It is an application of the magnetocaloric effect, first discovered in iron in 1883~\cite{Warburg1883}, whereby the temperature of a suitable material can be changed upon the application of a magnetic field. This occurs in materials that are paramagnetic by virtue of an electronic magnetic moment or as a result of the nuclear spin. Nuclear paramagnets are most relevant for the temperature range discussed in this review, and the corresponding cooling technique is known as adiabatic nuclear demagnetisation.

The principle of demagnetisation cooling is well established. For overviews, see for example~\cite{Lounasmaa1974,Pobell2007,Pickett2000}. Here we provide a brief outline for those unfamiliar with the topic to aid understanding of later sections. Nuclear demagnetisation refrigeration operates by controlling the Zeeman splitting of the nuclear spin energy levels in an applied magnetic field. For small magnetic fields, the Zeeman splitting is much less than the thermal energy $k_\mathrm{B}T$, leading to a random spin-orientation distribution throughout the refrigerant. This gives an entropy contribution of $S = R\ln (2I+1)$, for $R$ the ideal gas constant and $I$ the nuclear spin. In suitable materials~\cite{Pickett1988}, where this is the dominant entropy contribution, a significant entropy reduction can be obtained by ordering the spin orientations in a large magnetic field. This can be used as part of a cooling technique by first applying a magnetic field of $\sim 10\,\mathrm{T}$ and then waiting for the nuclear spins to thermalise to the base temperature of a dilution refrigerator (a process termed precooling). The refrigerant is then thermally isolated from the mixing chamber of the dilution refrigerator, allowing it to remain at approximately constant entropy, and the magnetic field is swept down, producing cooling.

The molar nuclear spin entropy is approximately~\cite{Enss2005}
\begin{equation}
	S \approx R\ln (2I+1) - \frac{\lambda_n}{2\mu_0}\left(\frac{B}{T_\mathrm{n}}\right)^2 \;.
	\label{eq:nuclear_entropy}
\end{equation}
This shows that the entropy is entirely a function of $B/T_\mathrm{n}$, meaning that the minimum attainable final temperature is given by $T_f = T_iB_f/B_i$, where $T_i$ is the initial nuclear temperature, and $B_i$ \& $B_f$ are the initial and final magnetic fields, respectively. Note that the total magnetic field consists of the externally applied field $B_\mathrm{ext}$ and the effective nuclear internal field $b$, which arises from the magnetic dipole interactions in the nuclei. These fields combine to give the total field $B = \sqrt{B^2_\mathrm{ext} + b^2}$.

Cooling by demagnetisation often uses elaborate refrigeration stages~\cite{Pickett2000} on state of the art, custom-built dilution refrigerators~\cite{Cousins1999}, or vibration isolated systems on commercial, cryogen-free dilution refrigerators~\cite{Batey2013,Todoshchenko2014}. While these systems can readily reach bulk electron temperatures $\sim 100\,\mathrm{\mu K}$, it is not straightforward to use them to cool a nanoelectronic sample to similar temperatures. In the remainder of this review, we will discuss recently developed techniques that can be used to overcome some of the challenges and effectively apply demagnetisation refrigeration to micro/nanoelectronic devices and samples.

\section{I\lowercase{mmersion cooling}}
\label{sec:immersion}

In the context of low temperature micro/nanoelectronic devices, immersion cooling means immersing parts of the experiment, including the device, in liquid helium to improve thermal contact with the coldest stage of a refrigerator. This coldest stage may be the liquid helium refrigerant inside the mixing chamber of a dilution refrigerator or the solid refrigerant of a demagnetisation refrigerator. The helium in the immersion cell may be either $^3\mathrm{He}$, $^4\mathrm{He}$, or a mixture of the two and, depending on the working temperature, may be in either the normal state or the superfluid state. Thermal contact to liquid in the immersion cell is often improved by the use of sintered metal-powder heat exchangers, which provide an extremely large solid/liquid contact area to counteract the boundary resistance at the solid/liquid interface~\cite{Harrison1979,Pobell2007}. For example, a sintered silver heat exchanger with a volume of a few cubic centimetres may have a contact surface area $\sim 10\,\mathrm{m^2}$~\cite{Bunkov1991}. In the context of the thermal model discussed above, immersion cooling can be used to ensure that the substrate, off-chip wiring and the sample environment (which contributes to $\dot{Q}_\mathrm{par}$) are all well thermalised at the base temperature of a refrigerator.

Immersion cooling has been used to thermalise micro/nanoelectronic devices to the base temperature of dilution refrigerators~\cite{Samkharadze2011,Samkharadze2015,Bradley2016,Knighton2018,Nicoli2019,Lane2020} and demagnetisation refrigerators~\cite{Pan1999,Xia2000,Huang2007,Vliet2018}. In all cases where a separate immersion cell is used, sintered metal powder heat exchangers are used to make thermal contact between the helium in the cell and the cold metal parts of the refrigerator. In many cases, sintered silver heat exchangers in the immersion cell are also used to make good thermal contact with the off-chip wiring~\cite{Pan1999,Xia2000,Huang2007,Samkharadze2011,Samkharadze2015,Bradley2016,Knighton2018,Vliet2018}. The aim is to cool on-chip electrons through electronic heat conduction, exploiting the $T^{-1}$ scaling of the electronic thermal resistance (Equ.~\ref{equ:R_WF}) in preference to the $T^{-4}$ scaling of the electron-phonon thermal resistance (Equ.~\ref{equ:Q_ep}). This approach is particularly effective for samples with a low electrical contact resistance, and optimising sample fabrication for lower resistances can produce lower base electron temperatures~\cite{Vliet2018}.

Despite significant efforts, electron temperatures reached with immersion cooling are rarely below $\approx 4\,\mathrm{mK}$. Pan, Xia, Tusi and co-workers \cite{Pan1999,Xia2000} found an electron temperature of $4\,\mathrm{mK}$ in a GaAs/AlGaAs 2D electron gas using a $^3$He immersion cell cooled by a $\mathrm{PrNi}_5$ nuclear demagnetisation stage. Some of the same authors have also reported temperature-dependent behaviour down to $0.5\,\mathrm{mK}$ in a similar experiment~\cite{Huang2007}. At Lancaster University, Bradley et al. \cite{Bradley2016} reached an electron temperature of $3.7\,\mathrm{mK}$ in an experiment where a Coulomb blockade thermometer (CBT) was placed in the mixing chamber of a dilution refrigerator, rather than in a separate immersion cell. A $^3\mathrm{He}$ immersion cell cooled by a copper nuclear demagnetisation stage has been used to reach an electron temperature below $2\,\mathrm{mK}$ in a 2D electron gas, as measured using current-noise-sensing thermometry~\cite{Vliet2018}. While successful, experiments of the type described above require custom-made or significantly customised refrigerators. Nicol\'{i} et al.~\cite{Nicoli2019} developed an immersion cell to reach an electron temperature of $6.7\,\mathrm{mK}$ in a gated, GaAs/AlGaAs quantum dot in a commercial, cryogen-free dilution refrigerator. While still technically challenging, this experiment did not require significant modification of the dilution refrigerator (for example, opening the mixing chamber) and no additional magnetic cooling stage was added. 

In principle, immersion cooling can be used to reach low-millikelvin electron temperatures in a commercially-available dilution refrigerator. However, if significantly lower temperatures are needed, it is also necessary to employ nuclear demagnetisation refrigeration. In the following section, we describe how demagnetisation cooling can be used to directly refrigerate off-chip wiring, potentially bypassing the need for an immersion cell.

\section{D\lowercase{emagnetisation refrigeration of electrical contacts}}
\label{sec:cont_demag}

In traditional microkelvin experiments, the measurement wiring is typically thermalised at the lowest temperatures on a single nuclear demagnetization stage by wrapping a long section of wiring but making thermal contact only through a thin layer of electrical (and thus thermal) insulation, preventing undesired electrical shorting. At temperatures below $10\,\mathrm{mK}$, or certainly below $1\,\mathrm{mK}$, this becomes prohibitively inefficient. In this section we summarize the results obtained using networks of demagnetisation stages, where each measurement lead passes through its own nuclear refrigerator (NR). This eliminates the weak cooling link through an electrical insulator and replaces it with electronic Wiedemann--Franz cooling. The approach has been implemented in three successive versions at the University of Basel. We describe these experimental setups and review measurements of micro/nanoelectronic devices cooled through nuclear refrigeration of their measurement leads. The first two generations of nuclear stages were developed for a Leiden cryogenics MNK wet dilution refrigerator. The third generation was installed on a Bluefors LD dry dilution refrigerator.

\begin{figure}
	\centering
	\includegraphics[width=0.9\columnwidth]{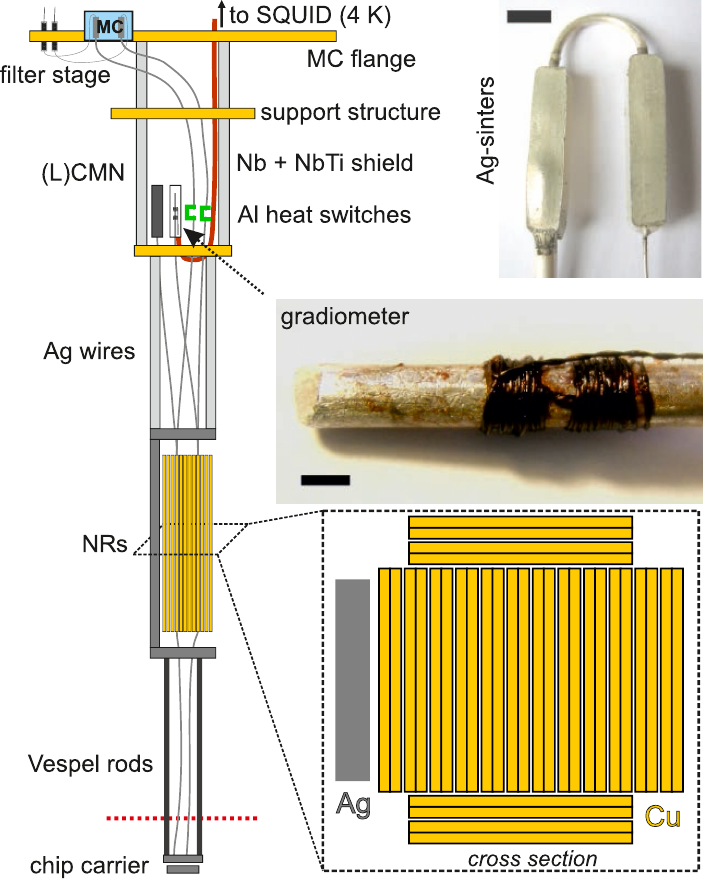}
	\caption{Schematic of a nuclear demagnetization stage mounted on a Bluefors LD dry dilution refrigerator. The measurement leads are thermalised with Ag powder sinters~(\emph{top right picture}, scale bar:~$5\,\mathrm{mm}$) in the mixing chamber~(MC, blue) and pass through C-shaped Al heat switches~(green) to the Cu plates. The gradiometer of a noise thermometer as well as the~(L)CMN thermometers are positioned in a region of cancelled magnetic field between the MC and the NR stage. The gradiometer is double-shielded by a Nb tube and a outer NbTi tube~(red). \emph{Middle right inset:}~photograph of the gradiometer pick-up coil made from insulated Nb wire with $100\,\mathrm{\mu m}$ diameter. The $2 \times 20$ turns are wound non-inductively on a high-purity silver wire which is spot-welded to a NR. Scale bar:~$2\,\mathrm{mm}$. \emph{Lower inset:}~schematic cross section through the network of 16~parallel NRs. Each NR is $2\,\mathrm{mol}$ of Cu ($99.99\,\mathrm{\%}$ Cu, low-$\mathrm{H_2}$ content~\cite{aurubius}, $\mathrm{RRR} \sim 500$) and consists of two half-plates, spot-welded together at the top and bottom. Each half-plate is of dimension $3.4\,\mathrm{cm} \times 0.17\,\mathrm{cm} \times 12\,\mathrm{cm}$. This figure was taken from~\cite{Palma2017}.}
	\label{fig:BFscheme}
\end{figure}

A full schematic of the latest (3rd generation) demagnetisation setup, installed on a Bluefors LD refrigerator, is shown in Fig.~\ref{fig:BFscheme}. With increasing generation of demagnetization stage, the lowest electron temperature after demagnetization in the NRs was reduced from $1\,\mathrm{mK}$ in~\cite{Clark2010} to $0.2\,\mathrm{mK}$ in~\cite{Feshchenko2015} and finally to $0.15\,\mathrm{mK}$ in~\cite{Palma2017}. The improvements result mainly from increasing the amount of copper per plate ($0.57\,\mathrm{mol}$\,/\,$1\,\mathrm{mol}$\,/\,$2\,\mathrm{mol}$) while optimizing the geometry for reduced eddy-current heating and increasing the diameter of the silver wires ($1.27\,\mathrm{mm}$\,/\,$1.27\,\mathrm{mm}$\,/\,$2.54\,\mathrm{mm}$) connecting the NRs to silver sintered heat exchanges residing inside the mixing chamber. Finally, the surface area of the silver sintered heat exchangers, as determined from BET surface area analysis~\cite{Brunauer1938}, was increased from $3\,\mathrm{m^2}$ in the first two generations to $9\,\mathrm{m^2}$ in the third generation. An overview of relevant system parameters for the different generations of demagnetization stage is given in table~\ref{tab:overviewBS}.

Measurement setups on both dilution fridges (wet and dry) use $\approx 1.5\,\mathrm{m}$ long thermocoax cables from room temperature down to the mixing chamber, which are excellent microwave filters in the few GHz to the high THz regime~\cite{Scheller2014,Zorin1995,Glattli1997}. The wires are thermally anchored at all relevant temperature stages of the dilution refrigerator. In order to also obtain strong microwave attenuation in the MHz regime, additional home-built Ag-epoxy microwave filters with $>100\,\mathrm{dB}$ attenuation for frequencies above $\approx200\,\mathrm{MHz}$ are installed at the MC level in both experimental setups. Transmission spectra for the microwave filters and a thermocoax cable for comparison are shown in Fig.~\ref{fig:filters}. The filters consist of $\approx2.5\,\mathrm{m}$ of Cu wire with thin insulation, embedded into a conductive Ag-epoxy matrix, thus leading to excellent thermalisation properties in addition to the filtering~\cite{Scheller2014}, as demonstrated on a wet dilution refrigerator without a demagnetization stage where electron temperatures of $7.5\,\mathrm{mK}$ were obtained in two metallic Coulomb Blockade Thermometers (CBTs).

Three different types of nanoelectronic devices have been investigated on the second generation nuclear stage: quantum dots fabricated on GaAs/AlGaAs heterostructures~\cite{Maradan2014}, normal metal-insulator-superconductor tunnel junctions (NIS)~\cite{Feshchenko2015}, and metallic CBTs~\cite{Casparis2012}. All devices were installed at the sample stage located some distance below the NRs and were held at constant magnetic field (zero field in case of the quantum dot and NIS samples and a small finite field in case of the CBTs). The nuclear demagnetization experiments discussed in the following therefore reveal information solely about cooling devices through their electrical contacts.

\begin{figure*}
	\centering
	\includegraphics[width=1.3\columnwidth]{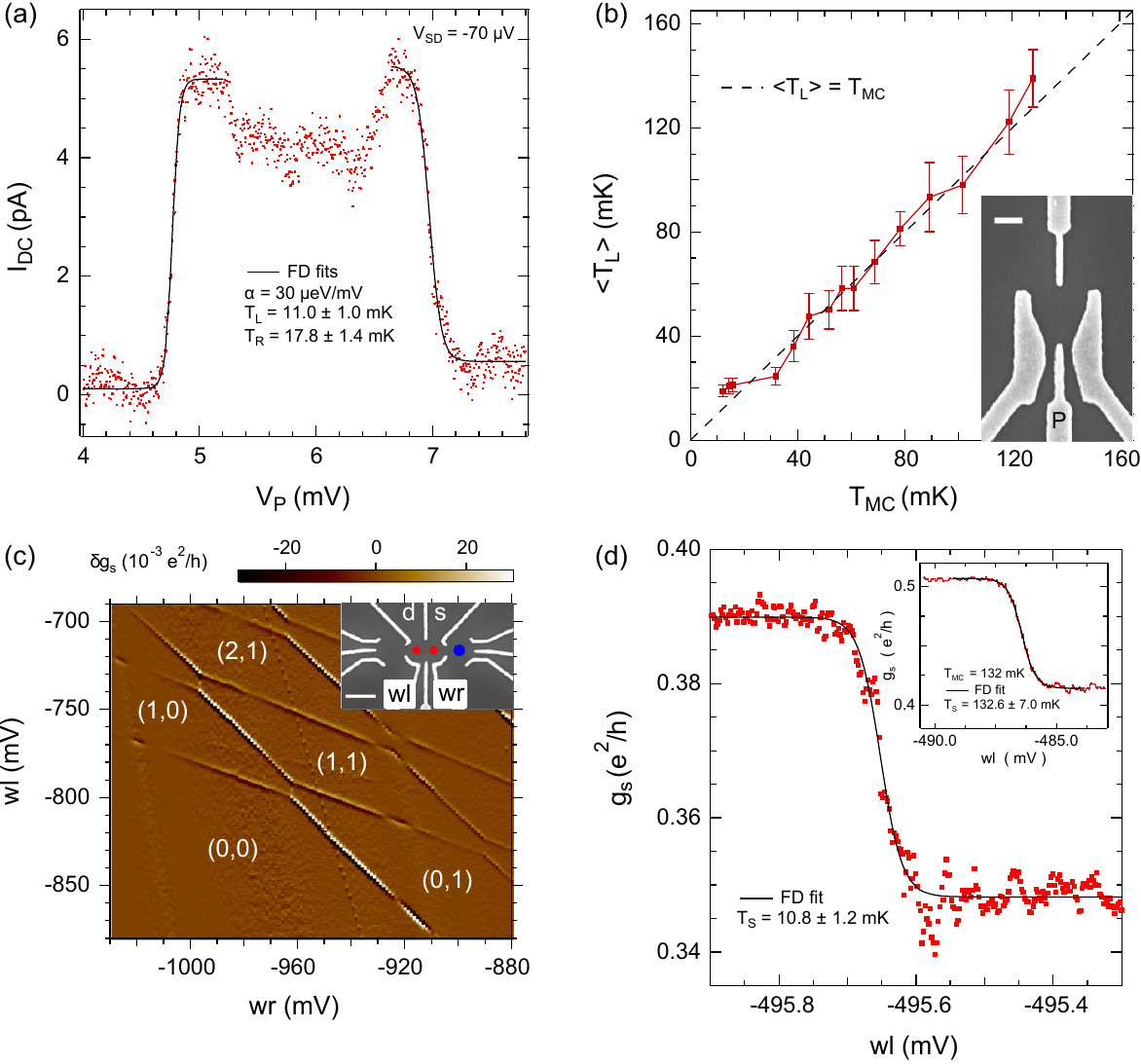}
	\caption{Electron temperature extracted from direct DC transport (a,b) and charge sensing (c,d) of quantum dots formed at the heterointerface of GaAs/AlGaAs structures. (a) Measured DC current (red dots) along with Fermi-Dirac curve fits (solid black curves). (b) Extracted electron temperature $T_L$ from measurements as shown in (a) as a function of mixing chamber temperature $T_{MC}$. The inset shows a SEM image of a similar device. (c) Charge stability diagram of a double quantum dot similar to the one shown in the inset. (d) Electron temperatures extracted by fitting a Fermi-Dirac distribution to the charge sensing signal at base temperature and $T_{MC}=132\,\mathrm{mK}$ are shown in the main panel and inset, respectively. This figure was adapted from~\cite{Maradan2014}.}
	\label{fig:QD}
\end{figure*}

The GaAs quantum dots were investigated in two modes of operation, direct transport and charge sensing. In the first method, a small source drain bias of $V_\mathrm{SD} = 70\,\mathrm{\mu V}$ was applied to a single quantum dot and the resulting DC current, shown in Fig.~\ref{fig:QD}, was measured as a function of plunger gate voltage $V_\mathrm{p}$ used to shift the quantum dot level with respect to the source and drain chemical potential. In the limit of small tunnelling rates, the temperature broadening of the resulting DC current steps can be fit with a Fermi-Dirac distribution to obtain separately the electronic temperature of the adjacent source and drain leads. Strictly speaking, this method holds only in two-dimensional systems where the density of states is constant and thus the shape of the current profile (given by the integral over energy of the density of states multiplied by their occupation probability) is determined only by the Fermi--Dirac distribution. In practice, changes in the density of states for 1D and 3D systems are small on the energy scale of a few $\mathrm{\mu eV}$ such that this analysis is also valid for current leads with any dimensionality (complications may arise from local mesoscopic fluctuations that can induce significant changes in the density of states). The resulting electron temperature obtained from direct DC transport was $11\pm 1\,\mathrm{mK}$ at a refrigerator temperature of $9\,\mathrm{mK}$.

The second method of quantum dot thermometry relies on charge sensing using the double quantum dot device shown in the inset of Fig.~\ref{fig:QD}(b) (SEM image of a similar device). The charge stability diagram obtained from the sensor quantum dot located on the right-hand-side of the device is shown in Fig.~\ref{fig:QD}(c) and exhibits the typical honeycomb structure for double quantum dots with charge occupation as indicated in the figure. Charge sensing works with with arbitratily low tunneling rates in the dots, allowing them to remain in the temperature broadened regime to aribtrarily low temperatures. Thermometry was carried out by scanning across the (0,0)--(0,1) charge transition line as a function of the right wall [`wr' in Fig.~\ref{fig:QD}(c)] gate voltage and fitting the current profile obtained from the sensor quantum dot with a Fermi-Dirac distribution.  The lowest electron temperature obtained from 6 consecutive charge sensing traces in this case was $10.3\pm4.4\,\mathrm{mK}$. These measurements were mainly limited by $1/f$ noise from the semiconductor wafer. 

We note that in contrast to direct transport measurements, where the lever-arm (the conversion between gate voltage and quantum dot energy) can be extracted directly from a single DC bias trace and is given by the width of the current step and applied DC bias, the present method requires high temperature calibration. Here the sample is heated up to a regime where the mixing chamber temperature and electronic device temperature are assumed to agree, which then allows for extraction of the lever-arm from the broadening of the charge transition line. Alternatively, the charge transition line may be also explored as a function of applied DC bias, see~\cite{Maradan2014}, resulting in a lever-arm of unity for the given set of parameters, thus eliminating the need for high temperature calibration. Consistent results have been obtained with both approaches. We note two difficulties in measuring ultra-low temperatures using quantum dots. First, due to the small dimensionality of the device, those systems are very susceptible to charge fluctuations in the host waver material, typically on the order of $1\,\mathrm{\mu eV}$~\cite{Camenzind2018}, and second, voltage noise in the electrical contacts can translate directly into an elevated electron temperature reading.

\begin{figure*}
	\centering
	\includegraphics[width=1.8\columnwidth]{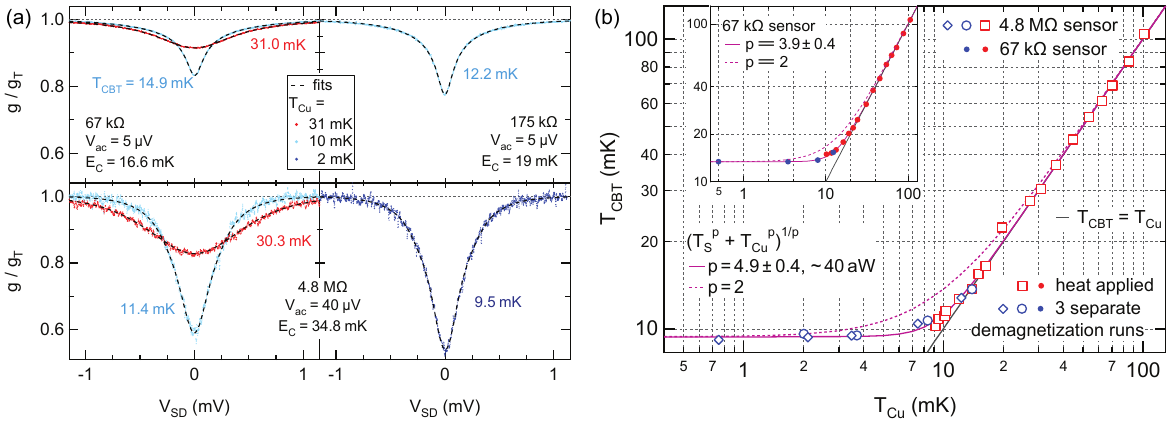}
	\caption{Thermometry using various metallic Coulomb blockade thermometers with differing resistance. Normalised differential conductance $g/g_T$ as a function of applied DC bias is shown in panel (a) for various copper plate temperatures $T_{Cu}$. Off-chip demagnetization down to $T_\mathrm{Cu}=2\,\mathrm{mK}$ slightly reduces the electronic temperature for the $4.8\,\mathrm{M\Omega}$ device from $11.4\,\mathrm{mK}$ (light blue) to $9.5\,\mathrm{mK}$ (dark blue). (b) Extracted electron temperatures as a function of $T_\mathrm{Cu}$. Open (closed) markers represent the $67\,\mathrm{k\Omega}$ ($4.8\,\mathrm{M\Omega}$) device. Data for the red and blue markers were collected during regular dilution refrigerator operation and adiabatic nuclear demagnetization, respectively. This figure was adapted from~\cite{Casparis2012}.}
	\label{fig:CBT_MNK}
\end{figure*}

Metallic CBTs are simple to use two-terminal devices that allow for precise thermometry down to the few millikelvin regime and below. The devices consist of parallel chains of metallic islands separated by tunnel junctions (usually aluminium oxide). In contrast to quantum dots, which are operated in deep Coulomb blockade, the CBT islands are in the high temperature limit where their charging energy $E_C$ is comparable to the thermal energy $k_\mathrm{B}T_\mathrm{e}$. The conductance of the CBT exhibits a dip around zero bias, and both the width and the depth of the dip are temperature dependent and can be used for thermometry. The applied AC and DC bias, and any voltage noise, is equally divided between the junctions in each chain of islands on the CBT. This reduces the demands on the environmental noise level compared to quantum dot thermometry. Fig.~\ref{fig:CBT_MNK}(a) shows the bias dependence of three CBTs with differing total resistance. A full fit to the bias dependence (dashed black curves) delivers the charging energy of the device and the corresponding electron temperature~\cite{Pekola1994}.

Higher electron temperatures ($14.9\,\mathrm{mK}$\,/\,$12.2\,\mathrm{mK}$\,/\,$11.4\,\mathrm{mK}$) were obtained for the lower resistance devices ($67\,\mathrm{k\Omega}$\,/\,$175\,\mathrm{k\Omega}$\,/\,$4.8\,\mathrm{M\Omega}$) during regular operation of the dilution refrigerator, consistent with the notion of better isolation from the environment due to larger resistances. Only little cooling is observed during the off-chip adiabatic nuclear demagnetization (the device resides on a sample holder at compensated magnetic field), lowering the electron temperature of the $4.8\,\mathrm{M\Omega}$ device from $11.4\,\mathrm{mK}$ to $9.5\,\mathrm{mK}$ upon reduction of the Cu-plate temperature from $\approx 10\,\mathrm{mK}$ to $2\,\mathrm{mK}$. This is not so surprising since Wiedemann--Franz cooling through the sample leads is expected to be effective only for low impedance devices, as illustrated in Fig.~\ref{fig:thermal_resistances}. Presumably the small temperature reduction upon demagnetization results mainly from the sample holder being cooled by a nuclear refrigerator through a massive, $99.999\%$ pure (5N), silver wire which results in slightly improved cooling of the CBT through its insulating substrate.

\begin{figure*}
	\centering
	\includegraphics[width=1.4\columnwidth]{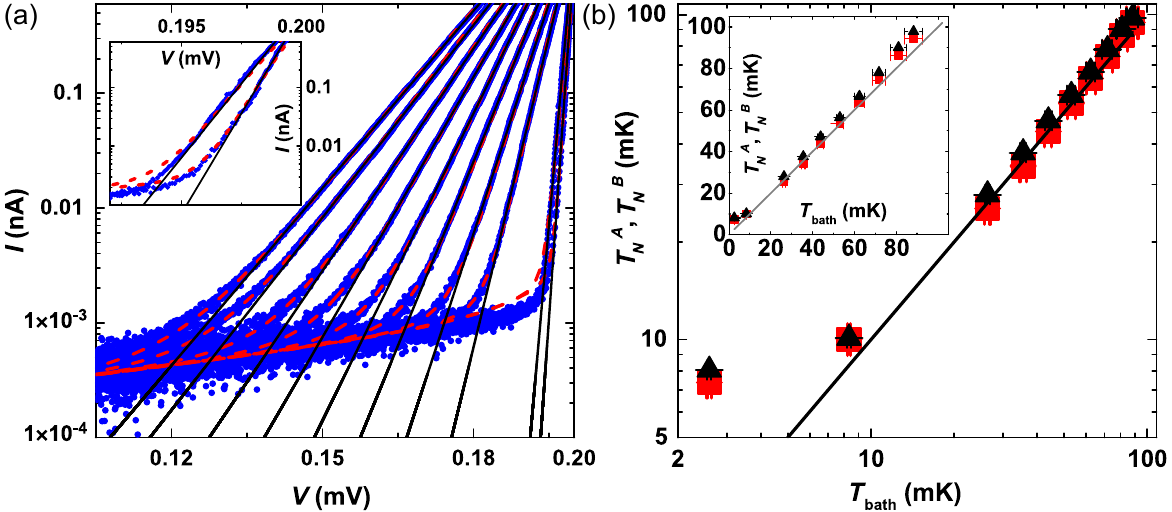}
	\caption{Normal metal-insulator-superconductor (NIS) tunnel junction thermometry. (a) Linear fits (solid black) to the onset of the measured quasiparticle current (blue dots) in an NIS device. Fits to the full current profile are shown in dashed red. The inset shows a close-up for mixing chamber (bath) temperatures of $10\,\mathrm{mK}$ and $7\,\mathrm{mK}$ on the left and right, respectively. (b) Extracted electronic temperatures from (a) for the full curve fit and the linear fit are shown as red squares and black triangles, respectively. This figure was adapted from~\cite{Feshchenko2015}.}
	\label{fig:NIS}
\end{figure*}

Next we review thermometry results from normal metal-insulator-superconductor (NIS) devices. The sharp quasiparticle peak in the density of states of the superconductor provides an ideal probe to measure the thermal broadening of the distribution of occupied states in the adjacent normal metal. In order to do so, an insulator is sandwiched between the normal metal and the superconductor. This is such that the creation of Cooper pairs is highly suppressed, and therefore the resulting DC current through the device, as a function of bias voltage $V$, reflects the superconducting gap. The electron temperature $T_N^A$ can then be directly extracted by performing a linear fit [solid black lines in Fig.~\ref{fig:NIS}(a)] to the onset of the quasiparticle current $I$ in logarithmic scale, i.e. $T_N^A=e/k_\mathrm{B}\cdot dV/d(\mathrm{ln}\,I)$ where $e$ and $k_\mathrm{B}$ are the elementary charge and the Boltzmann constant, respectively. Alternatively, a fit to the full bias profile can be applied [dashed red curves in Fig.~\ref{fig:NIS}(a)] to extract the electron temperature of the normal metal as described in detail in~\cite{Feshchenko2015}.

Due to the huge, mm-size macroscopic leads on the NIS device, one could hope for improved off-chip nuclear demagnetization performance compared to the high impedance arrays present in the metallic Coulomb blockade thermometers. Indeed, the electron temperature drops by $\approx30\,\%$ from \mbox{$\approx10\,\mathrm{mK}$} to \mbox{$\approx7\,\mathrm{mK}$} in Fig.~\ref{fig:NIS}(b) upon reducing the Cu-plate temperature down to $3\,\mathrm{mK}$, compared to only a $\approx15\,\%$ reduction in temperature in the case of the CBTs in Fig.~\ref{fig:CBT_MNK}(b). The limiting factor in this case is most likely the RMS voltage noise $\left< V_\mathrm{N}^2 \right>$ in the measurement leads which couples directly to the chemical potential in the normal metal and translates into an elevated temperature reading if $\left< V_\mathrm{N}^2 \right> \gg k_\mathrm{B}T_\mathrm{e}/e$. In addition, residual perpendicular magnetic fields also lead to a drastic overestimation of the electronic temperature \cite{Feshchenko2015}.

Table~\ref{tab:overviewBS} summarizes all the relevant system parameters such as sample mount, nuclear stage dimensions and mass, filtering, sinters and so forth for the three generations of nuclear stage installed on a wet MNK system from Leiden cryogenics (1st and 2nd generation) and on a dry LD system from Bluefors (3rd generation). In addition, an overview is given of the electron temperature measurements performed using quantum dots, NIS devices, and metallic CBTs. Details of the lowest temperature results, which were reached using CBTs on the 3rd generation stage, can be found in section~\ref{sec:5_Basel}.

\begin{table*}
	\begin{tabular}{|l||l|l|l|}
		\hline
		{\bf Generation stage}	            & {\bf 1st generation}~\cite{Clark2010}	& {\bf 2nd generation}~\cite{Casparis2012,Feshchenko2015}	& {\bf 3rd generation}~\cite{Palma2017}	\\
		\hline
		\hline
		Refrigerator model	          & Leiden cryogenics 	                  & Leiden cryogenics 	                           & Bluefors	\\
		                  	          & (MNK)            	                  & (MNK)            	                           & (LD)        	\\
		\hline
		Wet / dry system        	    & Wet               	                  & Wet	                                           & Dry	\\
		\hline
		Sample mount            	    & Sample stage       	                  & Sample stage                                   & Nuclear stage	\\
		\hline
		Demagnetization								& Off-chip                            & Off-chip                                     & On- \& off-chip\\
		\hline
		\# NRs	                      & $13$                        	          & $21$	  \                                         & $16$	\\
		\hline
		NR dimensions            	    & $10\cdot2\cdot0.2\,\mathrm{cm^3}$	          & $9\cdot3.2\cdot0.25\,\mathrm{cm^3}$	                 & $2\cdot(12\cdot3.4\cdot0.17\,\mathrm{cm^3}$)	\\
		\hline
		Cu NR mass	                  & $0.57\,\mathrm{mol}$                  	          & $1\,\mathrm{mol}$                         	               & $2\,\mathrm{mol}$	\\
		\hline
		Sinter surface area  	        & $3\,\mathrm{m^2}$	                            & $3\,\mathrm{m^2}$                                       & $2\cdot4.5\,\mathrm{m^2}$	\\
		\hline
		Ag wire diameter	            & $1.27\,\mathrm{mm}$	                            & $1.27\,\mathrm{mm}$                                       & $2.54\,\mathrm{mm}$	\\
		\hline
		Discrete filter @ MC	        & None                                  & \cite{Casparis2012}~RC 2-pole, $10\,\mathrm{kHz}$ BW	                           & RC 2-pole, $45\,\mathrm{kHz}$ BW	\\
		                    	        &                                       & $820\,\mathrm{\Omega}$/$22\,\mathrm{nF}$,  $1.2\,\mathrm{k\Omega}$/$4.7\,\mathrm{nF}$  & $2\cdot[2\,\mathrm{k\Omega}$/$680\,\mathrm{pF}$]	\\

		                    	        &                            & \cite{Feshchenko2015}~RC 2-pole, $30\,\mathrm{kHz}$ BW  &  	\\
		                    	        &                            & $1.6\,\mathrm{k\Omega}$/$2.2\,\mathrm{nF}$,  $2.4\,\mathrm{k\Omega}$/$470\,\mathrm{pF}$  &  	\\

		\hline
		Lowest NR $T_\mathrm{e}$    	& $1\,\mathrm{mK}$                                	& \cite{Casparis2012}~$0.3\,\mathrm{mK}$ / \cite{Feshchenko2015}~$0.2\,\mathrm{mK}$	                                       & $0.15\,\mathrm{mK}$	\\
		                              & Power curves	                        & Power curves                  	               & Noise thermometry	\\
		\hline
		Lowest sample $T_\mathrm{e}$	& Not measured	                        & QD~\cite{Maradan2014}: $10.4\,\mathrm{mK}$          	     & CBT~\cite{Palma2017a}: $2.8\,\mathrm{mK}$	\\
		                            	&                                     	& CBT~\cite{Casparis2012}: $9.5\,\mathrm{mK}$	               & ($1.8\,\mathrm{mK}$ in Fig.~\ref{fig:demagBilal})	\\
																	&                                     	& NIS~\cite{Feshchenko2015}: $7\,\mathrm{mK}$	               &  	\\
		\hline
	\end{tabular}
	\caption{Comparison of the three different Basel nuclear stages, the first two on the same wet system and the third on a dry system. For all systems $\approx1.5\,\mathrm{m}$ of uninterrupted thermocoax cable was used for the measurement wires down to the mixing chamber, at which different cold filters were mounted. For the 3rd generation, two half plates were spot welded together for the purpose of reducing eddy-current heating and two Ag sinters with $4.5\,\mathrm{m^2}$ surface area were installed for each measurement lead. The lowest electron temperatures for the nuclear stage and sample are indicated in the last two rows.}
	\label{tab:overviewBS}
\end{table*}

The experiments discussed above show that low-millikelvin on-chip electron temperatures can be successfully reached by magnetic refrigeration of external electrical connections. These experiments also demonstrate that the the base electron temperature is often limited by the device being measured, not the external refrigerator. In the case of quantum dots and NIS thermometers, intrinsic noise (charge fluctuations), extrinsic noise (voltage fluctuations) and residual perpendicular magnetic field (for the NIS thermometer) likely limited the lowest $T_\mathrm{e}$ that could be resolved. In the case of CBTs, their high impedance meant that cooling through electrical connections was less effective. In the following section, we discuss how on-chip magnetic refrigeration can be used to overcome the latter challenge.

\section{O\lowercase{n-chip demagnetisation refrigeration}}
\label{sec:chip_demag}

On-chip demagnetisation refrigeration uses a small quantity of refrigerant integrated onto a micro/nanoelectronic device. The refrigerant is electrically connected to the device's conduction electrons, providing a thermal link to the nuclear spins via hyperfine interactions between the nuclei and electrons \cite{Pobell2007,Gloos1991}. This bypasses the electron-phonon coupling bottleneck associated with cooling a sample through its electrically insulating substrate. It also bypasses the weak thermal link to off-chip wiring in high impedance devices.

The earliest observations of on-chip magnetic cooling were made where, instead of using a conventional nuclear demagnetisation refrigerant such as copper, the spin entropy was provided by electronic paramagnetism within the material of the device structure. In \cite{Bestwick2015}, which is an investigation into the anomalous Hall effect in a topological insulator, an unexpected variation in the Hall bar's resistivity was found and ultimately identified as the result of unexpected temperature changes. These temperature changes arose from a magnetocaloric effect in some unknown part of the device. During experiments, the device temperature was reduced to $25\,\mathrm{mK}$ from a mixing chamber temperature of $40\,\mathrm{mK}$. This resulted in a very low longitudinal resistance and excellent Hall conductance quantisation. Unexpected cooling has also been observed in measurements of aluminium SETs~\cite{Ciccarelli2016}. In this work, the aluminium was doped with manganese in order to suppress superconductivity, which was undesirable for good device operation. The doping was found to have the side effect of allowing demagnetisation refrigeration of the SET to $140\,\mathrm{mK}$, down from the $300\,\mathrm{mK}$ base temperature of the $\mathrm{^3He}$ cryostat in which the sample was mounted.

For on-chip cooling to a few millikelvin, the most effective approach to date uses relatively small blocks of metallic refrigerant in direct electrical connection with the circuit elements of a device. Provided the connection has a low enough electrical resistance, the conduction electrons in the device and the refrigerant are essentially a single thermal bath, cooled by demagnetisation of the refrigerant's nuclear spins. A number of demonstrations have been made using CBTs to measure electron temperature during the cooling process, at Lancaster University \cite{Bradley2017}, the University of Basel \cite{Palma2017a} and Delft University of Technology \cite{Yurttaguel2019,Sarsby2019}. The CBT is particularly well suited to the demonstration of magnetic cooling since the operation of the device itself is insensitive to the applied magnetic field \cite{Hirvi1995,Hirvi1996,Pekola1998}, and it can also be fabricated with conveniently sized metallic islands for the addition of refrigerant, which can be electroplated up to a thickness of $\sim 10\,\mathrm{\mu m}$ [see Fig.~\ref{fig:lancs_dry_demag}(a) and (b)]. Electroplating is used to avoid stress build-up in the thick metal film, which often occurs with more conventional deposition techniques (e.g.\@ sputtering or evaporation).

\subsection{Demagnetisation cooling with only on-chip refrigerant}

\begin{figure*}
	\includegraphics[width=1.2\columnwidth]{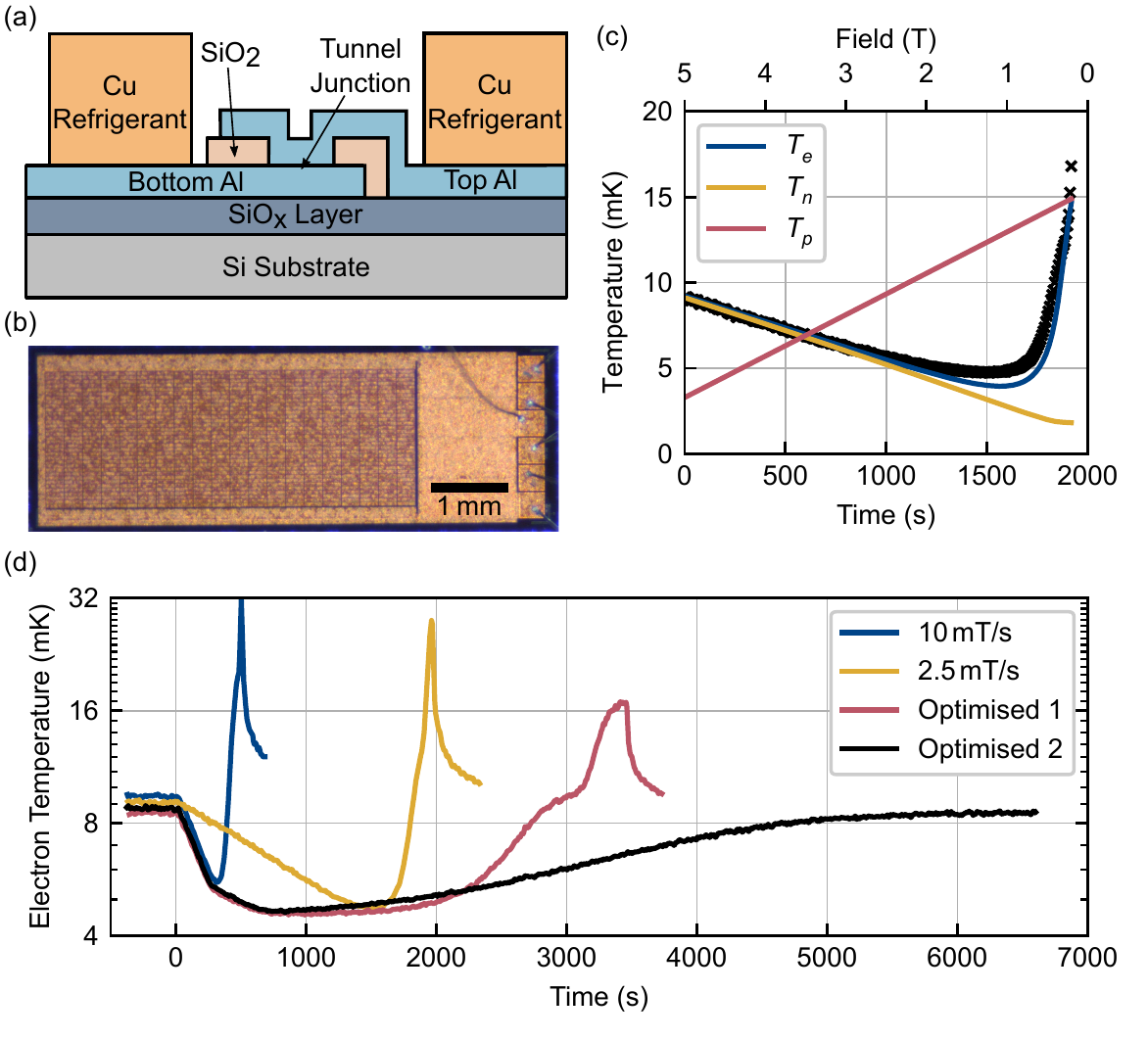}
	\caption{Demonstration of on-chip demagnetisation refrigeration with copper refrigerant. The CBT device shown schematically in (a) features large ($6\,\mathrm{\mu m}$ thick) $\mathrm{Cu}$ refrigerant blocks applied to an array of metallic islands. A photograph of the $6.5\,\mathrm{mm} \times 2.3\,\mathrm{mm}$ chip is shown in (b), with the $32 \times 20$ array of metal islands taking up the left $3/4$ of the device. The black crosses in panel (c) show the measured electron temperature during a $2.5\,\mathrm{mT/s}$ demagnetisation, to which the three subsystem model was fitted, allowing extraction of the phonon and nuclear spin temperatures. Panel (d) shows how the base temperature and hold time were extended by using three different demagnetisation rates instead of one. Details of the demagnetisation profiles `Optimised 1' and `Optimised 2' can be found in \cite{Bradley2017}.}
	\label{fig:lancs_dry_demag}
\end{figure*}

On-chip nuclear refrigeration was first demonstrated using $6\,\mathrm{\mu m}$ thick copper refrigerant electroplated onto the $32 \times 20$ metal island array of a CBT device \cite{Bradley2017}. This sample was pre-cooled to $T_\mathrm{e} \approx 9\,\mathrm{mK}$ using a cryogen-free dilution refrigerator, with a base temperature of $7\,\mathrm{mK}$, in a $5\,\mathrm{T}$ magnetic field. When demagnetising from $5\,\mathrm{T}$ at a rate of $2.5\,\mathrm{mT/s}$, the CBT conductance was seen to drop as would be expected for a falling on-chip electron temperature. Repeated experiments made with different DC biasing of the CBT confirmed that the conductance change was indeed due to a change in temperature, and not the result of electromagnetic induction. The lowest temperatures reached with such single-rate demagnetisations were $T_\mathrm{e} \approx 5\,\mathrm{mK}$, significantly below the base temperature of the dilution refrigerator. 

Electron temperature data from single-rate demagnetisation experiments were compared to predictions of the thermal model described in section~\ref{sec:heatflow}. The temperature of the nuclear spins was assumed to reduce adiabatically as the magnetic field was stepped down, with the electrons being cooled by heat flow to these spins in competition with the incoming heat via electron-phonon coupling and parasitic heating. The model was found to be consistent with the electron temperature data in Fig.~\ref{fig:lancs_dry_demag}(c) and a dynamic (during the sweep) heat leak of $6.3\,\mathrm{fW}$ per island, also confirming that the heat flow to the nuclear spins goes as $B^2$, as expected from Equ.~\ref{equ:Q_en}. With the heat leak due to eddy-current heating going as $(dB/dt)^2$~\cite{Pobell2007}, it was expected that reducing the ramp rate as the demagnetisation proceeded to lower fields would lead to lower base temperatures (see also~\cite{Gachechiladze1986,Strehlow2006}). The result of this optimisation is shown by the third (red) and fourth (black) traces in Fig.~\ref{fig:lancs_dry_demag}(d), in which the latter line shows the benefit of having a larger nuclear heat capacity if the demagnetisation is completely stopped at a higher magnetic field. Optimisation of the demagnetisation profile resulted in a slightly lower base electron temperature of $4.5\,\mathrm{mK}$ and a significantly longer hold time: around 1200\,s below $5\,\mathrm{mK}$.

\begin{figure*}
	\includegraphics[width=1.2\columnwidth]{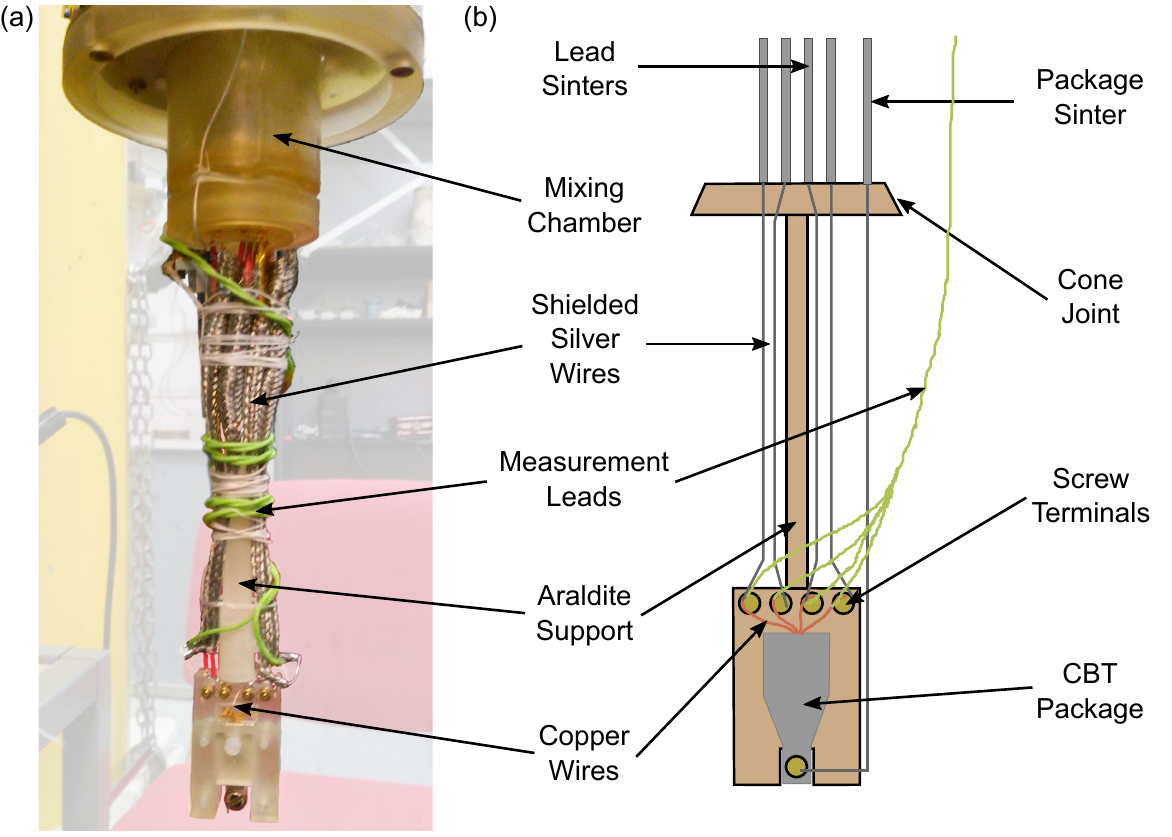}
	\caption{The `coldfinger' used for precooling a CBT sensor in its package on a dilution refrigerator in Lancaster. Panel (a) shows the coldfinger mated with the mixing chamber hence, when cooled, the sinters shown at the top of the diagram in panel (b) are immersed in the liquid $\mathrm{^3He}$--$\mathrm{^4He}$ refrigerant of the dilution refrigerator. Cooling is provided through the silver wire connected to the package and also the shielded silver wires attached to each measurement lead.}
	\label{fig:coldfinger}
\end{figure*}

\begin{figure*}
	\includegraphics[width=1.2\columnwidth]{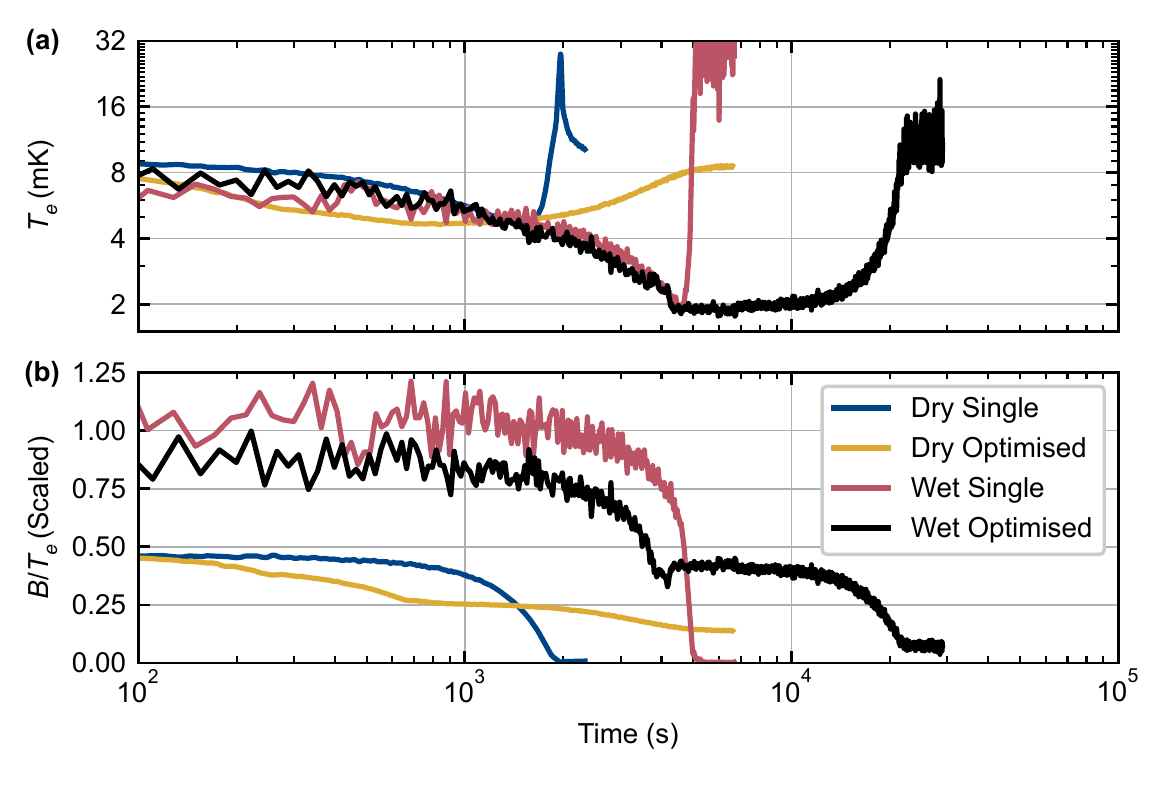}
	\caption{Comparison of demagnetisation cooling using the same cooling platform on wet and dry dilution refrigerators. Panel (a) shows a comparison of the electron temperatures achieved during single rate and optimised multi-rate demagnetisations on the wet and dry dilution refrigerators. Panel (b) shows a quantity related to the entropy change during the demagnetisations, and therefore shows the amount of deviation from the ideal case of constant entropy.}
	\label{fig:lancs_wet_demag}
\end{figure*}

As discussed in section~\ref{sec:heatflow}, the minimum possible temperature that can be reached during adiabatic demagnetisation is set by the initial entropy reduction achieved during magnetisation and precooling. As the entropy is given by Equ.~\ref{eq:nuclear_entropy}, we see that it is favourable to maximise the value of $B/T_\mathrm{n}$ by using larger magnetic fields and lower precooling temperatures. A similar CBT was therefore cooled in a different, Lancaster-built dilution refrigerator with an $8\,\mathrm{T}$ superconducting solenoid and a base temperature of $2.3\,\mathrm{mK}$~\cite{Bradley1994}, offering a potential five-fold improvement in $B/T_\mathrm{n}$ over the dry cryostat. The Lancaster-built cryostat features an openable plastic mixing chamber \cite{Frossati1978} and sintered silver heat exchangers were added to the mixing chamber to help precool the CBT, as shown in Fig.~\ref{fig:coldfinger}(a) and (b). This cryostat also has the inherent benefit of lower mechanical vibrations because there is no pulse-tube cooler, from which~\cite{Wit2019} there can be a significant additional heat leak through eddy-current heating~\cite{Todoshchenko2014,Palma2017} and additional electrical noise~\cite{Kalra2016}. The particular dilution refrigerator used for the results shown in Fig.~\ref{fig:lancs_wet_demag} also features extensive vibration isolation and is located within a shielded room which further removes vibrations and electrical noise.

In Fig.~\ref{fig:lancs_wet_demag}(a), we see that the transition to a colder dilution refrigerator significantly improved the base electron temperature from $4.7\,\mathrm{mK}$ to $2.0\,\mathrm{mK}$ for the unoptimised single-rate demagnetisations, and from $4.5\,\mathrm{mK}$ to $1.9\,\mathrm{mK}$ for the optimised multi-rate demagnetisations. The latter case, where a significant magnetic field was held following the demagnetisation in order to maintain significant nuclear heat capacity, also shows a much increased hold time of some $6000\,\mathrm{s}$ around $2\,\mathrm{mK}$. These improvements are further reflected in Fig.~\ref{fig:lancs_wet_demag}(b) which shows the quantity $B/T_\mathrm{e}$, scaled such that its initial value is equal to unity at the start of the wet demagnetisations. As described in section~\ref{sec:heatflow}, the entropy of the system is entirely a function of $B/T_\mathrm{n}$, and since the electron-phonon coupling here is extremely weak compared to the electron-nuclear spin coupling, we can assume $T_\mathrm{e} = T_\mathrm{n}$. Figure~\ref{fig:lancs_wet_demag}(b) therefore shows deviation from the ideal case of constant entropy, which would be represented by a straight horizontal line. There is a clear initial benefit to the use of a cryostat with a lower base temperature and higher field magnet, since this leads to a larger initial $B/T_\mathrm{e}$ value and hence a larger entropy reduction during precooling. We also see that the optimised sweeps are able to avoid the sudden entropy change as the nuclear heat capacity is exhausted at the end of the single-rate sweeps.

Traditionally, the naturally abundant copper isotopes $\mathrm{^{63}Cu}$ and $\mathrm{^{65}Cu}$, both with spin $I = 3/2$, have been used for large bulk demagnetisation stages capable of themselves reaching electron temperatures of $12\,\mathrm{\mu K}$~\cite{Gloos1988} and cooling liquid helium to $100\,\mathrm{\mu K}$~\cite{Tuoriniemi2002}. Copper has been widely used both due to its thermodynamic benefits, such as a relatively large nuclear magnetic moment for all isotopes and low temperature of spontaneous magnetic ordering, but also more practical considerations such as the ease at which it can be machined into a desired shape and its good availability in high purity form~\cite{Pickett2000,Pobell2007}. However, there are other materials which have some benefits over copper, particularly in terms of the magnitude of the Korringa constant which determines the thermal coupling between the nuclear spins and conduction electrons. 

An alternative nuclear refrigerant is indium, which has spin $I = 9/2$, nuclear Curie constant $\lambda_n / \mu_0 = 13.8\,\mathrm{\mu J K T^{-2} mol^{-1}}$ and Korringa constant $\kappa = 0.09\,\mathrm{Ks}$. Indium therefore seems promising when compared to copper, which has smaller $\lambda_n / \mu_0 = 3.22\,\mathrm{\mu J K T^{-2} mol^{-1}}$, and hence a smaller nuclear heat capacity, and longer Korringa constant $\kappa = 1.2\,\mathrm{Ks}$~\cite{Pobell2007}, meaning weaker electron-nuclear spin coupling. Yet indium is mechanically soft, features an electric quadrupole interaction, which causes nuclear orientation below $300\,\mathrm{\mu K}$~\cite{Tang1985}, and has a superconducting transition at $28\,\mathrm{mT}$~\cite{Eisenstein1954}, limiting the lowest temperatures that can be reached during demagnetisations. This means indium has seldom been used for the construction of bulk demagnetisation stages. However, for on-chip cooling, where the refrigerant is applied by electroplating, and the minimum temperatures obtained are currently above $300\,\mathrm{\mu K}$, these limitations are not necessarily important.

Yurttag\"{u}l et al. at Delft University of Technology have demonstrated on-chip magnetic cooling using $25\,\mathrm{\mu m}$ thick, on-chip indium refrigerant blocks \cite{Yurttaguel2019}. These blocks were electroplated onto a CBT consisting of a $35 \times 15$ array of metallic islands. Precooling was performed using a `wet' dilution refrigerator equipped with a $12.8\,\mathrm{T}$ magnet and reached an initial electron temperature of $16\,\mathrm{mK}$. Following a demagnetisation at a rate of $0.4\,\mathrm{mT/s}$, a minimum electron temperature of $3.2\,\mathrm{mK}$ was obtained at a field of $2\,\mathrm{T}$, followed by rapid warming to above the initial electron temperature when the magnetic field ramp was stopped at $40\,\mathrm{mT}$, similar to what was observed for the unoptimised field sweeps in the copper experiments.

For both the indium and copper on-chip demagnetisations, the minimum electron temperature was found to be heavily influenced by the heat leak into the electrons on each of the CBT islands. This is particularly important since the CBT islands were permanently linked to the mixing chamber of the dilution refrigerator via the electron-phonon coupling and conduction through the measurement leads, with no controllable heat switch to break this link during the demagnetisation. While this makes for easy construction of the cooling platform, a penalty is paid in terms of the continuous heat input, particularly from phonons, when the CBT electrons are cooled significantly below the temperature of the fridge. Therefore, one approach for improving the minimum electron temperatures is to thermally isolate the device using a heat switch \cite{Bradley1984} and to cool the environment surrounding the CBT chip. This has been performed by combining the on-chip demagnetisations with demagnetisation of both the incoming measurement lines and the box the sample is mounted in, as described below.

\subsection{Magnetic cooling with on-chip and off-chip refrigerant}\label{sec:5_Basel}

Coulomb blockade thermometers with on-chip copper refrigerant have been studied in Basel using the 3rd generation magnetic refrigeration stage on a Bluefors LD dilution refrigerator (see section~\ref{sec:cont_demag} and table~\ref{tab:overviewBS} for details). In this case, the heat-leak into the cold on-chip islands is reduced by ensuring that substrate phonons and the off-chip wiring are also cooled below the base temperature of the dilution refrigerator.

\begin{figure*}
	\centering
	\includegraphics[width=1.6\columnwidth]{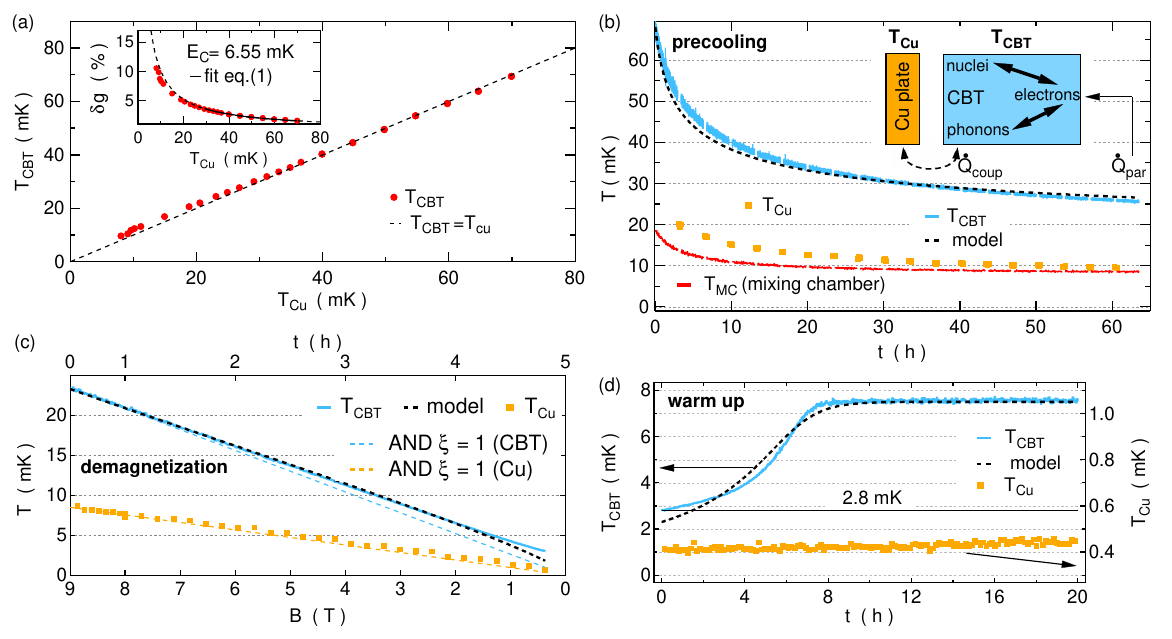}
	\caption{Nuclear adiabatic demagnetization of a metallic Coulomb blockade thermometer. (a) CBT temperature $T_\mathrm{CBT}$ versus copper plate temperature $T_\mathrm{Cu}$. The diagonal dashed line indicates ideal thermalisation $T_\mathrm{CBT}=T_\mathrm{Cu}$. The inset shows the normalized zero bias conductance dip $\delta g$ as a function of $T_\mathrm{Cu}$. A fit \cite{Meschke2016} (solid black curve) in the high temperature regime $T_\mathrm{Cu}>30\,$mK is used to extract the charging energy $E_\mathrm{C}=6.55\,$mK. The three steps of nuclear demagnetization, precooling, demagnetization, and warmup, are shown in panels (b), (c), and (d) respectively. Light blue, yellow, and red data indicate $T_\mathrm{CBT}$, $T_\mathrm{Cu}$, and $T_\mathrm{MC}$, respectively. Black dashed curves in (b-d) are predictions from a thermal model schematically indicated in (b). A light blue line in (c) indicates ideal adiabatic demagnetization. This figure was adapted from \cite{Palma2017a}.}
	\label{fig:CBT_BF}
\end{figure*}

\begin{figure*}
	\centering
	\includegraphics[width=1.2\columnwidth]{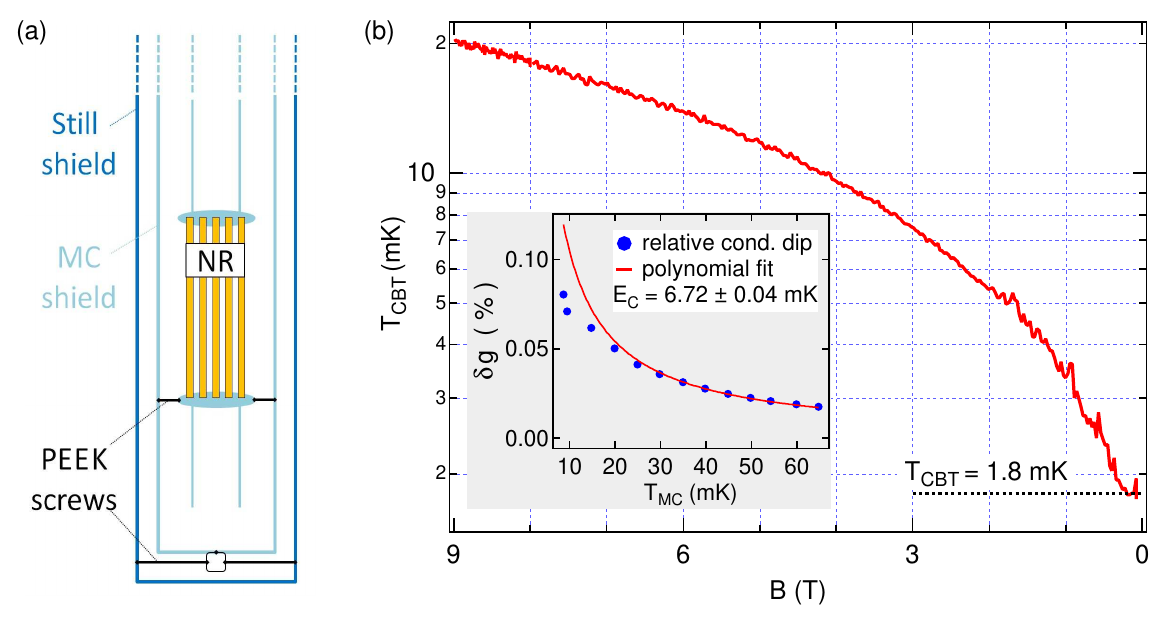}
	\caption{Nuclear adiabatic demagnetization of a metallic Coulomb blockade thermometer. (a) Schematic showing the demagnetization stage being fixed rigidly with respect to mixing chamber shield. A second set of PEEK screws fixes the mixing chamber shield with respect to the still radiation shield. (b) The main panel shows the extracted CBT temperature as a function of magnetic field during the demagnetization process. Calibration data are shown in the inset, where blue markers correspond to measurements of the relative conductance dip $\delta g/g_T$ and a fit to the data in the high temperature regime from $30\,\mathrm{mK}$ to $65\,\mathrm{mK}$ is shown in solid red. The resulting charging energy is $E_C=6.72\pm0.04\,\mathrm{mK}$.}
	\label{fig:demagBilal}
\end{figure*}

While dry dilution refrigerators, such as the Bluefors LD, seem to be the future path of low temperature physics, with obvious advantages compared to wet systems such as lower operating costs and independence of the worlds helium production, there are also disadvantages. Stronger magnets are available for wet systems due to the more efficient cooling when immersing the magnet directly into liquid helium. Furthermore, the pulse tube coolers used in dry systems introduce higher levels of vibrations, which is detrimental for adiabatic nuclear demagnetization experiments due to vibration induced eddy current heating. It is these vibrations that result in relatively high CBT precooling temperatures in these experiments, as shown in Figs.~\ref{fig:CBT_BF},\ref{fig:demagBilal}. This is the current bottleneck for this setup.

In contrast to previous experiments using the Basel refrigeration stages, here the sample is placed inside a small copper box which is mounted directly onto a nuclear refrigerator while using two other NRs as sample leads. This allows for direct on-chip demagnetization of the copper electroplated CBT islands in addition to off-chip demagnetization. The CBTs are operated in secondary mode, i.e. recording only the zero bias conductance during demagnetization. While this method requires high temperature calibration it comes with the advantage that no DC current passes through the device which otherwise would lead to Joule heating effects. In fact, a single bias trace after demagnetization is sufficient to destroy the nuclear polarization in the Cu-plated CBT islands that was built up during precooling at large magnetic field. The Joule heating effect is already visible at the lowest temperatures obtained in continuous mode operation of the dilution refrigerator without demagnetization, as demonstrated in \cite{Scheller2014,Palma2017a}. 

The inset in Fig.~\ref{fig:CBT_BF}(a) shows the relative conductance dip size $\delta g=1-g(V_\mathrm{SD})/g_\mathrm{T}$ as a function of Cu-plate temperature, where $g(V_\mathrm{SD})$ is the differential conductance as a function of applied source-drain bias $V_\mathrm{SD}$ and $g_\mathrm{T}$ the temperature-independent high-bias differential conductance. The relative conductance dip size can be approximated by $\delta g=u/6-u^2/60+u^3/630$ where $u=E_\mathrm{C}/(k_\mathrm{B} T_\mathrm{CBT})$ \cite{Meschke2016}. Therefore a fit to $\delta g$ in the high temperature regime where $T_\mathrm{CBT}=T_\mathrm{Cu}$ allows one to extract the charging energy $E_C$ as the only free fit parameter. Subsequently, any measured conductance dip can be converted back to an electronic temperature $T_\mathrm{CBT}$ using the previously determined charging energy. The CBT agrees very well with the Cu-plate temperature and only starts to deviate slightly at low temperatures, reaching $T_\mathrm{CBT}=9.7\,$mK at $T_\mathrm{Cu}=8.1\,$mK. Here, for the third generation nuclear stage the Cu-plate temperature is determined directly by noise thermometry. In order to do so a gradiometer (non-inductive coil with 20 clockwise turns following 20 counter clockwise turns) is used to measure the magnetic noise created by the Brownian motion of electrons within a massive 5N silver wire spot-welded to a NR, see Fig.~\ref{fig:BFscheme}. The superconducting wires from the gradiometer are then fed through concentric Nb and NbTi shields up to the 4K stage where a superconducting quantum interference device (SQUID) is used to amplify the small voltage fluctuations. See \cite{Palma2017} for more details on this noise thermometry setup. 

When precooling the device at a large magnetic field in Fig.~\ref{fig:CBT_BF}(b), the CBT reaches a temperature of 24\,mK after more than 60\,h precooling time, significantly higher than the mixing chamber temperature $T_\mathrm{MC}$ and Cu-plate temperature $T_\mathrm{Cu}$, which both saturate just below 10\,mK. The high precooling temperature is limited by pulse tube vibrations leading to either eddy current induced heating in the CBT islands and/or voltage fluctuations in the measurement wires that are then dissipated through Joule heating in the sample. The pulse tube vibrations are clearly visible in voltage noise measurements across the device (see supplemental information in \cite{Palma2017a}) showing up as frequency combs with a $1.4\,\mathrm{Hz}$ spacing in-between peaks. In the subsequent demagnetization step in Fig.~\ref{fig:CBT_BF}(c) the CBT temperature drops by a factor of 8.6, reaching $2.8\,\mathrm{mK}$ at the end of the adiabatic demagnetization. After completing the demagnetization, the CBT immediately starts to warm up in Fig.~\ref{fig:CBT_BF}(d), reaching equilibrium at $T_\mathrm{CBT}=7.5\,\mathrm{mK}$ after roughly $8\,\mathrm{h}$ while the external NRs remain at microkelvin temperatures. This highlights the importance of on-chip demagnetization for metallic Coulomb blockade thermometers, and the thermal isolation of the on-chip islands from the off-chip wiring.

In an initial attempt to reduce parasitic heating caused by the pulse tube vibrations, additional fixing mechanisms were introduced, shown in Fig.~\ref{fig:demagBilal}(a). Insulating screws made from PEEK (polyether ether ketone) were used to attach the support structure of the parallel network of nuclear stages with respect to the mixing chamber radiation shield, and a second set of PEEK screws fixed the still radiation shield with respect to the mixing chamber shield.

The CBT device investigated here is nominally the same as in Fig.~\ref{fig:CBT_BF}, but mounted perpendicular to the demagnetisation field in contrast to the measurements shown in Fig.\,\ref{fig:CBT_BF}. The high temperature calibration required for operating the device as a secondary thermometer is shown in the inset of Fig.~\ref{fig:demagBilal}(b), giving a charging energy of $E_C=6.72 \pm 0.04\,\mathrm{mK}$. Fixing the NRs with respect to MC and still shields, together with an increased precooling time of $140\,\mathrm{h}$ results in a CBT temperature of $20.3\,\mathrm{mK}$ at the beginning of the adiabatic nuclear demagnetization. In addition, compared to the results in Fig.~\ref{fig:CBT_BF}, the ratio of initial and final electron temperature increased from $8.6$ to $11.3$, giving a final CBT temperature of $T_\mathrm{CBT}=1.8\,\mathrm{mK}$ in the Bluefors LD system~\cite{Kalyoncu2019}.

On- and off-chip demagnetisation refrigeration have also been combined by Sarsby et al. at Delft University of Technology, but with indium as the refrigerant \cite{Sarsby2019}. They employed indium refrigerant blocks electroplated on the islands of a CBT, similar to that used for the on-chip indium investigation \cite{Yurttaguel2019}, but with each of the four electrical measurement lines also passing through a macroscopic indium block in the fridge. Both the CBT and the lead NRs were precooled in a `wet' dilution refrigerator in a magnetic field of $12\,\mathrm{T}$ over a period of approximately $7\,\mathrm{days}$, giving a starting temperature of $13\,\mathrm{mK}$. After this, the authors employed a continuously variable demagnetisation rate, proportional to magnetic field, in order to balance the available nuclear cooling power against the eddy current heating. When the demagnetisation ended, at a final field of $100\,\mathrm{mT}$, the CBT electron temperature was $420\,\mathrm{\mu K}$. The electron temperature then remained below $700\,\mathrm{\mu K}$ for some $85\,\mathrm{hours}$, owing to a small heat leak of $27\,\mathrm{aW}$ per island.

\section{C\lowercase{onclusions and open questions}}
\label{sec:conc}
 
Techniques for cooling micro/nanoelectronic devices to ultralow temperatures have progressed significantly in the last five years, largely through the development of new experimental methods based on nuclear demagnetisation refrigeration. By using multiple, macroscopic demagnetisation refrigerators to cool the substrate and electrical contacts of a device, and by incorporating microscopic volumes of nuclear refrigerant into a device structure, it is now possible to produce and measure low- and sub-millikelvin electron temperatures on-chip. The continuing development of immersion cells cooled by demagnetisation refrigeration may also provide a solution, particularly for very low impedance devices. Despite these advances, the sensitivity of on-chip electrons to parasitic heating and electrical noise mean that it is still experimentally challenging to get the electrons cold and to perform accurate thermometry. Coulomb blockade thermometers have proven to be an excellent testbed for new cooling techniques, as they provide both reliable thermometry and a degree of built-in protection against electrical noise. It is an open question how these new cooling techniques can be applied effectively to other types of device. Based on the work to-date, it seems unlikely that one single approach to cooling will be effective for every type of micro/nanoelectronic device or sample. It also seems inevitable that careful consideration and design of the on-chip thermal environment will be needed for any experiment where sub-millikelvin electron temperatures are required.

\section*{A\lowercase{cknowledgements}}
This work was supported by the European Microkelvin Platform (the European Union's Horizon 2020 research and innovation programme, grant agreement No. 824109), the Swiss Nanoscience Institute, NCCR QSIT, Swiss NSF No. 179024 and an ERC starting grant (DMZ). The data used for Fig.~\ref{fig:lancs_wet_demag} and Fig.~\ref{fig:demagBilal} are available at \url{https://doi.org/10.5281/zenodo.3759633}. Data from the other figures was previously published, please see the respective references.

\raggedright
\bibliography{ult_nanoelectronic_devices}

\end{document}